\documentclass[prb,twocolumn,showpacs,amsmath,amssymb]{revtex4}

\usepackage{graphicx}
\usepackage{dcolumn}
\usepackage{bm}

\begin{document}

\preprint{APS/123-QED}

\title{Interplay of magnetic ordering and CEF anisotropy\\in the hexagonal compounds RPtIn, R = Y, Gd - Lu}

\author{E. Morosan, S. L. Bud'ko and P. C. Canfield}
\affiliation{Ames Laboratory and Department of Physics and
Astronomy, \\Iowa State University, Ames, IA 50011, USA}

\date{\today}

\begin{abstract}
Single crystals of RPtIn, $R~=$ Y, Gd - Lu were grown out of
In-rich ternary solution. Powder X-ray diffraction data on all of
these compounds were consistent with the hexagonal ZrNiAl-type
structure (space group P $\overline{6}$ 2 m). The $R~=$ Tb and Tm
members of the series appear to order antiferromagnetically
($T_N~=$ 46.0 K, and 3.0 K respectively), whereas the $R~=$ Gd, Dy
- Er compounds have at least a ferromagnetic component of the
magnetization along the c-axis. The magnetic ordering temperatures
of all of these systems seem to scale well with the de Gennes
factor dG, whereas the curious switching from ferromagnetic to
antiferromagnetic ordering across the series is correlated with a
change in anisotropy, such that, in the low temperature
paramagnetic state, $\chi_{ab}~>~\chi_c$ for the antiferromagnetic
compounds, and $\chi_c~>~\chi_{ab}$ for the rest. In order to
characterize the magnetic ordering across the RPtIn series, a
three-dimensional model of the magnetic moments in Fe$_2$P-type
systems was developed, using the \textit{three co-planar
Ising-like systems model} previously introduced for the extremely
planar TbPtIn compound: given the orthorhombic point symmetry of
the R sites, we assumed the magnetic moments to be confined to six
non-planar easy axes, whose in-plane projections are rotate by
$60^0$ with respect to each other. Such a model is consistent with
the reduced high-field magnetization values observed for the RPtIn
compounds, R$~=$ Tb - Tm, and qualitatively reproduces the
features of the angular dependent magnetization of
Ho$_x$Y$_{1-x}$PtIn at $H~=$ 55 kG.

\end{abstract}


\maketitle

\section{\label{sec:intro}I\lowercase{ntroduction}}

The RPtIn compounds (R = Y, La-Sm, Gd-Lu) have been reported to
crystallize in the ZrNiAl hexagonal
structure\cite{ferro01,zar02,gal03}, space group
\textit{P$\overline{6}$2m}, with the rare earth in orthorhombic
point symmetry, whereas the R = Eu member of this series forms in
an orthorhombic TiNiSi-type structure\cite{pot04,mul12} (space
group \textit{Pnma}). Magnetic and transport measurements on some
of these materials revealed a variety of physical properties
across the series: CePtIn\cite{fuj05,kur06,yam07,sat08} and
YbPtIn\cite{tro09,kac10,yos11} appear to be dense Kondo systems,
with the electronic specific heat coefficient $\gamma$ larger than
500 mJ/mol K$^2$, and 430 mJ/mol K$^2$ respectively; no magnetic
order was observed in the former compound down to 60
mK\cite{sat08}, whereas the latter appears to order
antiferromagnetically below 3.4 K\cite{tro09}. In
TbPtIn\cite{wat13} the antiferromagnetic nature of the magnetic
order below 50 K was indirectly suggested by the metamagnetic
transitions observed in the M(H) data below this temperature.
Watson \textit{et al.}\cite{wat13} have also reported that the R =
Gd and Dy members of this series have ferromagnetic ground states,
with T$_C$ = 89 K, and 38 K respectively, with reduced values of
both the effective and the saturated moments of these two
compounds. Whereas for the Dy system, the disagreement with the
respective theoretical values could be attributed to crystal
electric field CEF effects, it was unclear what was causing it in
GdPtIn. CEF effects are also apparent in the magnetization
measurements on PrPtIn down to T = 1.7 K, which, together with the
resistivity data\cite{zar02} suggest a possible ferromagnetic
transition at lower temperatures. Similar data on SmPtIn are
indicative of ferromagnetic ordering in this compound below
T$_C~=$ 25 K.

We recently presented detailed magnetization and transport
measurements on single crystals of TbPtIn\cite{mor14}: anisotropic
low-field susceptibility and specific heat measurements confirm
the antiferromagnetic ground state, with T$_N$ = 46 K, slightly
different than the previously reported value\cite{wat13}; below
the ordering temperature, complex metamagnetism is revealed by
magnetization measurements with applied field in the basal plane.
Whereas from the high-temperature inverse susceptibility we
obtained an effective moment $\mu_{eff}~=~9.74~\mu_{B}$/Tb$^{3+}$,
close to the theoretical value $9.72~\mu_{B}$, the high-field
magnetization data yielded values only up to
$\sim6~\mu_{B}$/Tb$^{3+}$, much smaller than the theoretical
saturated moment of $~9~\mu_{B}$. In order to explain the reduced
magnetization values, as well as the angular dependence of the
metamagnetic properties, we proposed a \textit{three co-planar
Ising-like systems model}, which took into account the
orthorhombic point symmetry of the rare earth ions in the
hexagonal unit cell of the RPtIn compounds\cite{mor14}. Within
such a model, for applied magnetic fields far smaller than the CEF
splitting energy, for TbPtIn one expects a crystal field limited
saturated paramagnetic CL-SPM state equal to $6~\mu_B$.

In view of these existing magnetization and transport data, the
magnetic ordering in the heavy rare earth members of the RPtIn
series was somewhat intriguing: whereas for GdPtIn and DyPtIn,
ferromagnetic ground states were reported, the intermediate R = Tb
member appears to order antiferromagnetically, which is a rather
unusual discontinuity for a magnetically ordering local moment
series.

In the present work we are trying to address this problem, and
also extend the characterization of the physical properties to all
the heavy RPtIn systems (\textit{i.e.}, for R = Y, Gd - Lu).
Having been able to grow single crystals for all of these
compounds, we have the possibility of determining the effect of
the CEF anisotropy on their magnetic properties, more so than in
the previous studies on polycrystalline samples. As we shall see,
the hexagonal crystal structure of these compounds, with three R
ions in the unit cell occupying unique orthorhombic point symmetry
sites, is of crucial importance in explaining the moment
configurations and magnetic ordering in the RPtIn materials
described here.

In presenting our data, we will start with the non-magnetic
members of the RPtIn series, R = Y and Lu; then we will continue
with the magnetic ones (R = Gd - Tm), characterizing each of the
compounds by temperature- and field-dependent magnetization, as
well as zero-field specific heat measurements. A brief description
of the previously reported\cite{tro09,kac10,yos11} heavy fermion
compound YbPtIn is also included, our measurements being
performed, as with all the other R compounds, on solution-grown
single crystals.

Next, we will briefly present the model for the magnetic moment
configuration, characterizing the extremely planar TbPtIn compound
(which is described in detail by Morosan \textit{et
al.}\cite{mor14}); a more generalized version of this model,
extended to three dimensions, will then be used to describe the
magnetism in the other magnetic compounds except YbPtIn.

We will conclude by emphasizing the possibility of generalizing
our \textit{three co-planar Ising-like systems model} to a three
dimensional one, which could potentially describe hexagonal
systems with orthorhombic point symmetry of the rare earth site
beyond the RPtIn series. Directions for further experiments which
could potentially probe the validity of our assumed model will
also be discussed.

\section{Experimental methods}

Single crystals of RPtIn (R = Y, Gd - Lu) were grown out of
high-temperature ternary solution\cite{mor14,fisk15,can16,can17}.
Initial concentrations used were typically R$_x$Pt$_x$In$_{1-2x}$,
with $x~=~0.05~-~0.10$, and the resulting crystals were
well-formed hexagonal rods. The In-rich self-flux was used because
it lowers the liquid-solidus surface of primary solidification for
RPtIn, and also introduces no new elements into the melt. After
placing the constituent elements in alumina crucibles, the
crucibles were sealed in quartz ampoules under partial argon
pressure. In most cases the ampoules were initially heated up to
$\sim~1200^0$ C, and then slowly cooled down to $\sim~800^0$ C,
over 50 to 100 hours. Subsequently, the excess liquid solution was
decanted, and the resulting hexagonal rods were, if necessary,
quickly etched in HCl to remove residual flux from the surface.
Higher decanting temperatures (\textit{i.e.}, above $1000^0$ C)
were necessary for R = Y, Gd and Er, whereas in the case of TmPtIn
the temperature interval for which best crystals were obtained was
lower (between $1100^0$ C to $750^0$ C). In some cases, the
hexagonal rods had hollow channels in the center, sometimes with
flux inclusions. We succeeded in optimizing the growth of TbPtIn
by using faster cooling rates (\textit{i.e.}, $400^{0}$ C / 50h);
this yielded only well-formed, full, hexagonal rods, whereas when
slowing down the cooling process hollow crystals were obtained
together with full, smaller ones. Slight modifications of the
initial concentrations and/or growth profiles for Dy, Ho or Er
didn't totally eliminate the formation of hollow rods, but
$100~\%$ dense samples could easily be found. For all our
measurements, carefully chosen, well-formed single crystals were
used, so as to avoid possible mass errors on hollow rods to
propagate into our data.

A drastic modification of the growth procedure needed to be made
for YbPtIn: a ternary solution with initial composition
Yb$_{0.4}$Pt$_{0.1}$In$_{0.5}$ was sealed in a 3-cap Ta
crucible\cite{mor21}, and slow-cooled from $1200^0$ C to $1000^0$
C over $\sim~100$ hours, resulting in well-formed hexagonal rods.

\begin{figure}
\begin{center}
\includegraphics[angle=0,width=120mm]{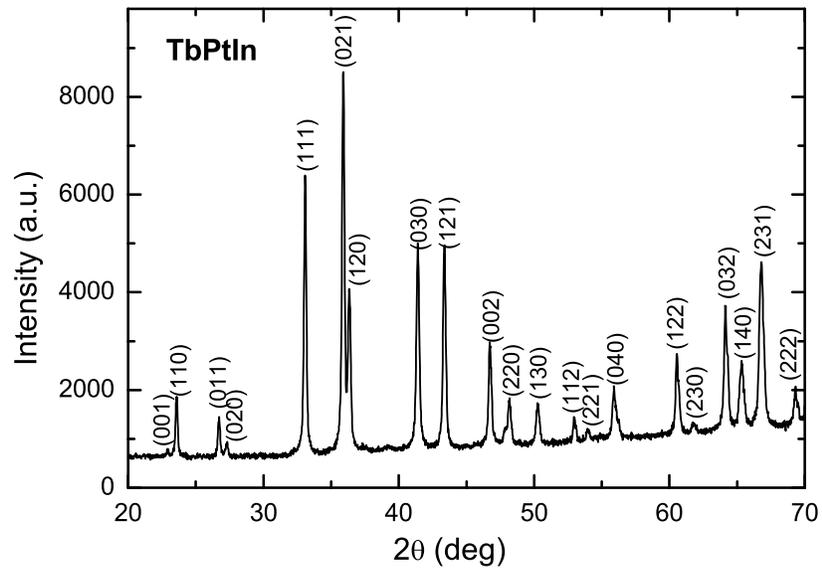}
\end{center}
\caption{Powder X-ray diffraction pattern for TbPtIn. All peaks
are indexed using a hexagonal P$\overline{6}$2m structure, with
$a~=~7.56~\AA$ and $c~=~3.87~\AA$.}\label{F02}
\end{figure}

\begin{figure}
\begin{center}
\includegraphics[angle=0,width=120mm]{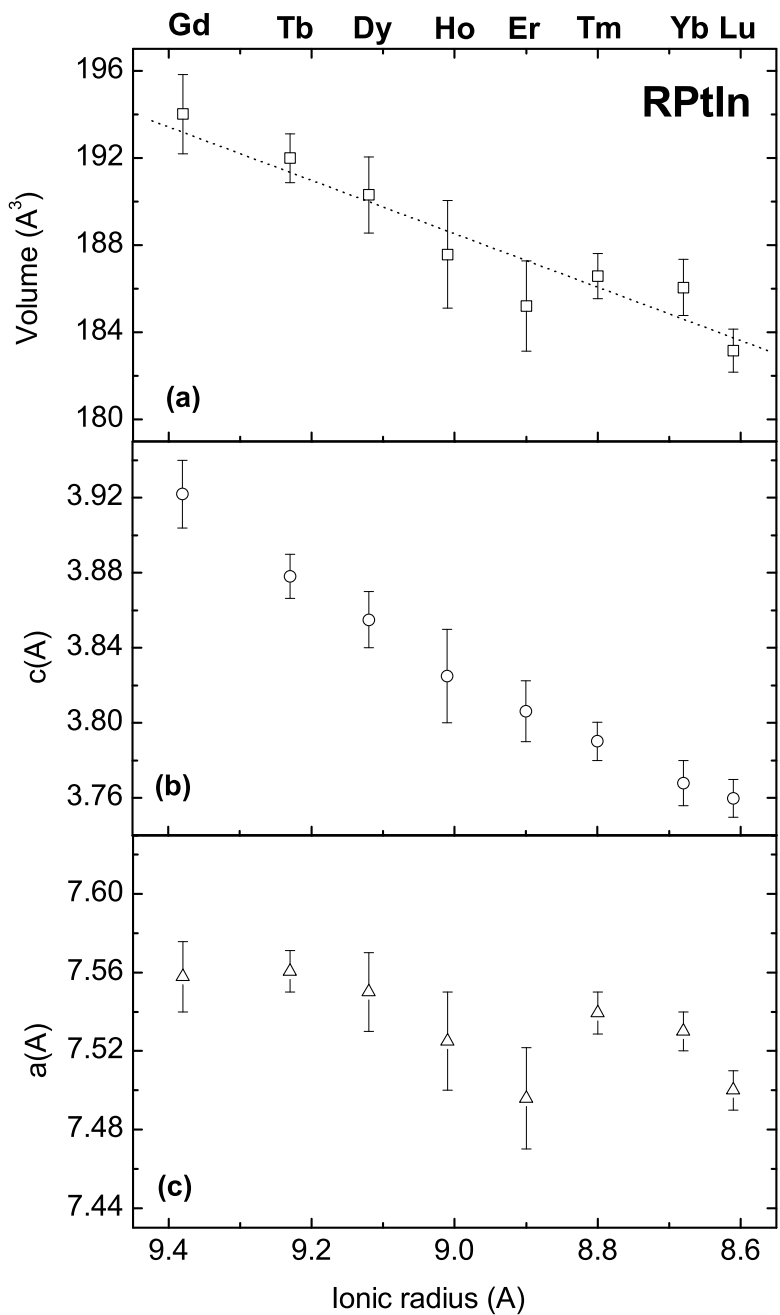}
\end{center}
\caption{Unit cell volumes and lattice parameters for RPtIn, R =
Gd - Lu, as a function of R$^{3+}$ ionic radius. Dotted line - a
guide for the eye, indicating the Lanthanide
contraction.}\label{F03}
\end{figure}

To confirm the crystal structure of the RPtIn compounds (R = Y, Gd
- Lu), room temperature powder X-ray measurements were performed,
using Cu K$_{\alpha}$ radiation. A typical pattern is shown in
Fig.\ref{F02} for TbPtIn. All detected peaks were indexed using
the Rietica Rietveld refinement program, with the
P$\overline{6}$2m space group and lattice parameters
$a~=~(7.56~\pm~0.01)~\AA$ and $c~=~(3.87~\pm~0.01)~\AA$, and no
secondary phases were detected. In Fig.\ref{F03}, the volume and
the dimensions of the unit cell across the series are shown as a
function of the R$^{3+}$ ionic radii. All values shown in this
figure are calculated from high angle/high intensity peaks,
\textit{i.e.}, using the (231) and (032) peaks, together with the
corresponding error bars which also encompass the values yielded
by the Rietica refinement. The unit cell volume (Fig.\ref{F03}a)
generally follows the expected lanthanide contraction\cite{sha18}
(as shown by the dotted line), as does the \textit{c} lattice
parameter (Fig.\ref{F03}b), whereas an apparent non-monotonic
change of the \textit{a} parameter is noticeable in
Fig.\ref{F03}c. However, the R = Tm and Yb volumes appear to
deviate slightly from the monotonic decrease across the series;
whereas this could indicate, at least for R = Yb, a trend toward
valence $2+$ for the rare earth ions, we shall see that the
magnetic measurements are consistent with the presence of
magnetism in the respective compounds, associated with trivalent R
ions.

Additional single crystal x-ray measurements were performed on the
R = Tb, Tm and Yb members of the series; these data indicated a
small (\textit{i.e.}, less than $6~\%$) deficient occupancy of one
of the two Pt sites in the unit cell of the R = Yb system, leading
to a stoichiometry closest to YbPt$_{0.97}$In. The analogous
measurements on TbPtIn and TmPtIn crystals indicated full
occupancy on all crystallographic sites in these compounds. The
lattice parameters from single crystal X-ray data for all three
compounds are consistent with the powder values in Fig.\ref{F03}.

Magnetic measurements were performed in a Quantum Design Magnetic
Properties Measurement System (MPMS) SQUID magnetometer (T = 1.8 -
350 K, H$_{max}$ = 55 kG). Additional measurements up to 140 kG
were also taken for TbPtIn, using an extraction magnetometer in a
Quantum Design Physical Properties Measurement System (PPMS). We
measured anisotropic field and temperature dependent magnetization
for all compounds, having the applied field $H~\parallel~c$ or
$H~\parallel~ab$, with arbitrary orientation of the field within
the basal plane unless otherwise specified. The corresponding
susceptibilities were calculated as $M~/~H$, whereas the
polycrystalline average susceptibility was estimated to be

$\chi_{ave}~=~1/3~*~(\chi_c~+~2~*~\chi_{ab})$,

or, when in-plane anisotropic measurements were available
(\textit{i.e.}, for the external field $H~\parallel~[100]$ and
$H~\parallel~[120]$),

$\chi_{ave}~=~1/3~*~(\chi_{[100]}~+~\chi_{[120]}~+~\chi_{[001]})$.

Curie-Weiss behavior of the anisotropic susceptibilities was
observed for all magnetic compounds, such that
$\chi~=~C~/~(T~+~\Theta_W)$. Thus from the high-temperature linear
inverse susceptibilities we were able to determine both the
effective moment $\mu_{eff}$ (and compare it with the expected
free ion value) and the anisotropic Weiss temperatures
$\Theta_{W}$ for each compound.

Additional angular dependent magnetization measurements have been
performed on the HoPtIn compound, as well as on the corresponding
dilution with non-magnetic Y$^{3+}$ ions on the Ho site,
Ho$_x$Y$_{1-x}$PtIn (x$~\simeq~0.04$). For these measurements, the
angular position of the samples was controlled by a specially
modified MPMS sample holder which allowed for the rotation of the
sample so that either the [001], [100] or [120]-axis stayed
perpendicular to the applied magnetic field. Torque on the rotor
was avoided by using small mass samples for the rotation
measurements, and these data were subsequently calibrated with the
two directions M(H) data on larger mass samples, as described in
more detail by Morosan \textit{et al.}\cite{mor14}.

Heat capacity measurements were made in the same Quantum Design
PPMS system. For each measurement, the sample holder and grease
background data, taken separately, were later subtracted from the
sample response.

For the antiferromagnetic members of the series (R = Tb and Tm),
we inferred the transition temperatures (N\'{e}el temperature or
spin reorientation temperatures) as determined from
$d(M_{ave}/H*T)/dT$ \cite{fis19} and $C_p(T)$ plots. In the
ferromagnetic RPtIn compounds, unusually broad peaks marked the
ordering in the specific heat data. Thus the onset of the
transition was chosen as the Curie temperature $T_C$
(\textit{i.e.}, the T value for which an increase of the specific
heat data occurs as the temperature is lowered through the phase
transition).

\section{Results}

We are characterizing each compound by anisotropic magnetization
and specific heat measurements, starting with the non-magnetic
YPtIn and LuPtIn members of the series. Next the magnetic RPtIn
will be introduced, for R = Gd to Tm. For each, we will emphasize
the nature of the ordered state together with the ordering
temperatures, as well as the high field, anisotropic magnetization
data, as these provide key values in our discussion and analysis
following the data presentation. Lastly, similar data on YbPtIn is
presented, with a note that a more detailed analysis of the heavy
fermion character of this compound is the subject of a separate
investigation\cite{mor21}.

\subsection{YPtIn and LuPtIn}

The anisotropic susceptibilities of the two members of the RPtIn
series with non-magnetic R ion (R = Y and Lu) are very small and
basically temperature-independent. However, the dominant terms in
the susceptibility data seem quite different for the two
compounds, as the average high-temperature values are positive in
the case of YPtIn (around $(6.6~\pm~0.7)~\times~10^{-5}$ emu/mol
for $H~\parallel~ab$, and $(4.6~\pm~1.1)~\times~10^{-5}$ emu/mol
for $H~\parallel~c$), and negative for LuPtIn (around
$(-3.8~\pm~0.2)~\times~10^{-4}$ emu/mol for $H~\parallel~ab$, and
$(-4.5~\pm~0.3)~\times~10^{-4}$ emu/mol for $H~\parallel~c$). The
field-dependent magnetization values for both compounds are
extremely small, as expected for non-magnetic R compounds.

\begin{figure}
\begin{center}
\includegraphics[angle=0,width=120mm]{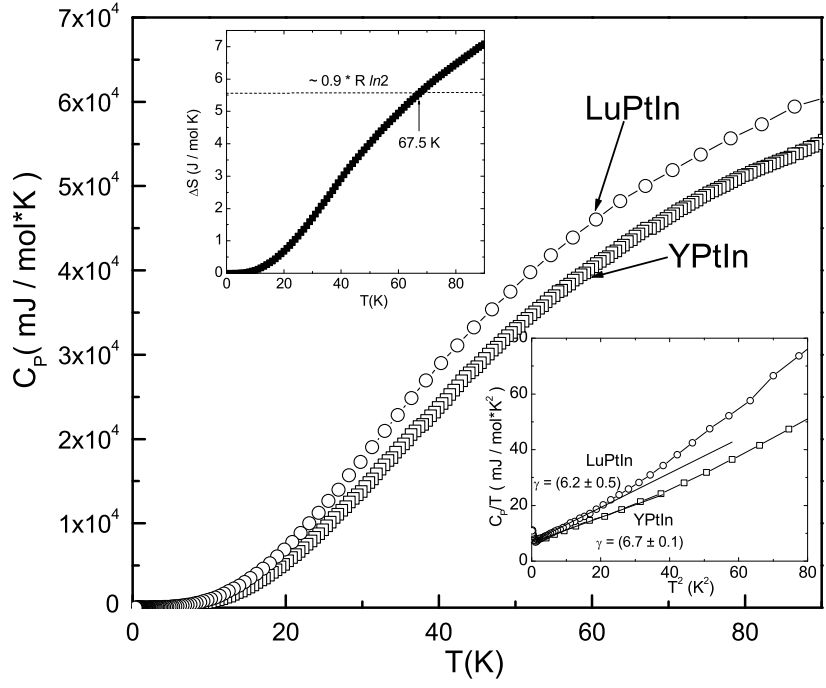}
\end{center}
\caption{Heat capacity for YPtIn and LuPtIn, with low-temperature
C$_P~/~T$ vs. T$^2$ shown in lower inset (linear fits of the low-T
data give $\gamma$ in units of mJ / mol * K$^2$); upper inset:
entropy difference $\Delta$S (see text).}\label{F04}
\end{figure}

Heat capacity measurements in zero applied field were performed
for the two systems, for $2~\leq~T~\leq~90$ K. As seen in Fig.
\ref{F04}, they have similar temperature dependencies; the higher
molecular weight for LuPtIn (and consequently the expected lower
Debye temperature) could explain the values of the specific heat
data being larger for this compound than for YPtIn. However, this
does not account for the big entropy difference for these two
systems (upper inset). This estimated entropy difference has a
fairly large value $\Delta~ S~\approx~0.9~*~R~\ln2$ around 67 K,
which, as we shall see, is the magnetic ordering temperature for
GdPtIn. (The entropy difference is still considerably large when
the measured specific heat data are scaled by their molecular
weights, according to the Debye model.) Consequently, no
meaningful magnetic specific heat estimates can be made for the
magnetic RPtIn compounds using either the R = Y or Lu compounds as
the non-magnetic analogues.

\subsection{RPtIn, R = Gd - Tm}

\textbf{GdPtIn}\\

\begin{figure}
\begin{center}
\includegraphics[angle=0,width=120mm]{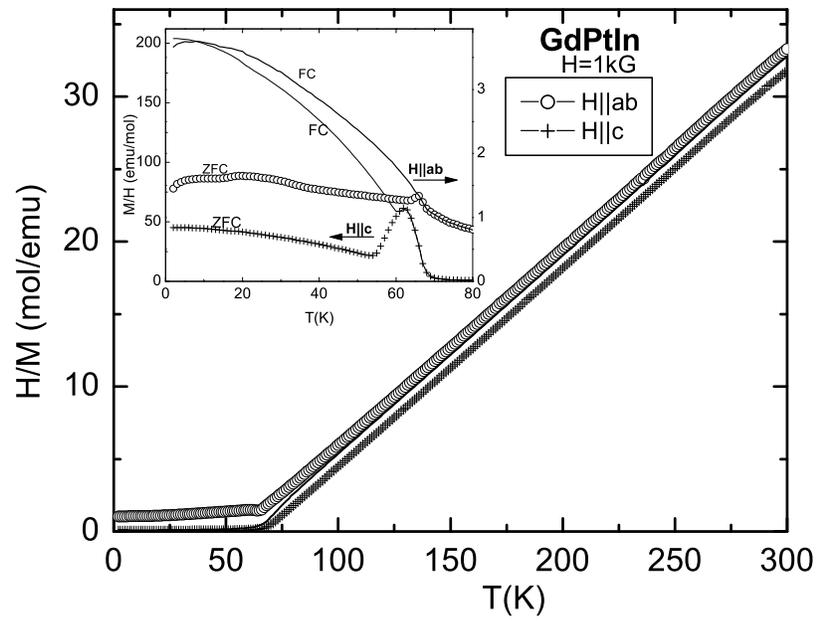}
\end{center}
\caption{Anisotropic $H~/~M$ data of GdPtIn and calculated average
(line) at H = 1 kG, with the anisotropic ZFC - FC low-temperature
$M~/~H$ data for H = 0.1 kG shown in the inset.}\label{F05}
\end{figure}

The anisotropic $H~/~M$ data for GdPtIn, together with the
polycrystalline average, are shown in Fig.\ref{F05}. The inset
presents the low-temperature $M~/~H$ data for low applied field (H
$=~100$ G), measured on warming up after either zero-field cooling
ZFC (symbols) or field cooling FC (solid lines) of the sample. The
paramagnetic susceptibility shows Curie-Weiss behavior
$\chi(T)~=~C/(T~+~\Theta_{W})$ above $\sim100$ K. $\Theta_{W}$
represents the Weiss temperature, which can be determined from the
linear fit of the high-temperature inverse susceptibility, and the
values corresponding to the two orientations of the field, as well
as the one for the polycrystalline average, are listed in Table 1.
The inverse susceptibility data appear slightly anisotropic,
contrary to the expected isotropic paramagnetic state for a Gd
compound. This is possibly caused by the dominant anisotropic
interaction in a pure S-state (\textit{i.e.}, L = 0)
compound\cite{jen}, also consistent with the different anisotropic
Weiss temperatures for this Gd system (see Table 1). The effective
moment value determined from the linear region in the inverse
average susceptibility is $\mu_{eff}~=~7.62~\mu_B$, comparable to
the theoretical value of $7.94~\mu_B$ for the Gd$^{3+}$ ions.

\begin{figure}
\begin{center}
\includegraphics[angle=0,width=120mm]{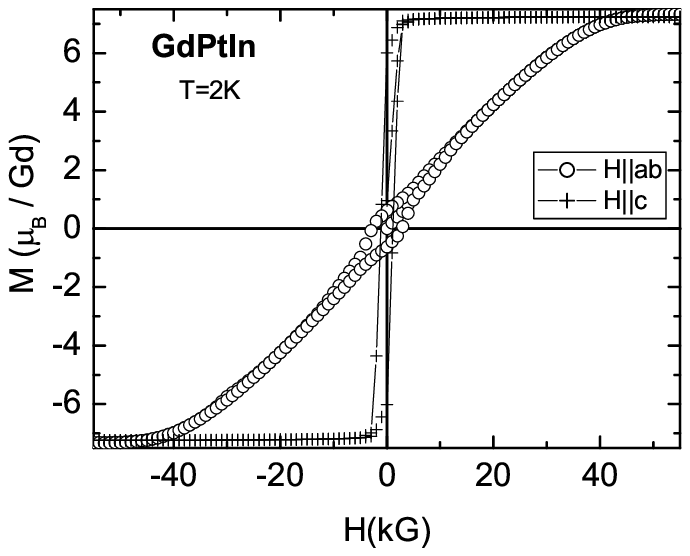}
\end{center}
\caption{Anisotropic field-dependent magnetization loops for
GdPtIn, at T = 2 K.}\label{F07}
\end{figure}

The field dependent magnetization data (Fig.\ref{F07}) appears to
indicate ferromagnetic ordering along the \textit{c}-axis.
Measurements performed for both increasing and decreasing applied
field reveal hysteresis loops for both H parallel and
perpendicular to the \textit{c}-axis. At $H~\approx~50$ kG, the
magnetization saturates in both directions around the expected
$7~\mu_B/$ Gd$^{3+}$ value. As the increase of the axial
magnetization with field (\textit{i.e.}, for $H~\parallel~c$) is
much faster than for $H~\perp~c$, we are led to believe that the
ferromagnetic exchange interaction favors moments' alignment along
the \textit{c}-axis.

\begin{figure}
\begin{center}
\includegraphics[angle=0,width=120mm]{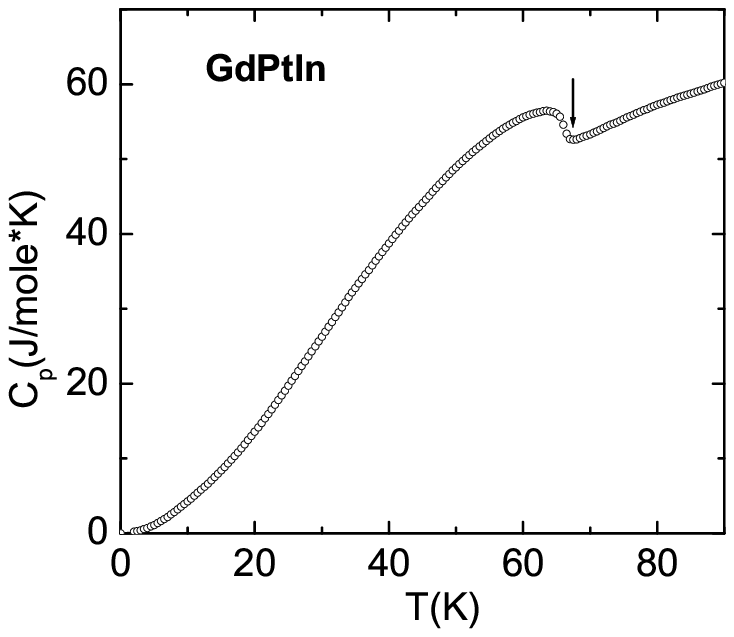}
\end{center}
\caption{Specific heat C$_P(T)$ of GdPtIn; small arrow indicates
T$_C$ determined from on-set (see text).}\label{F06}
\end{figure}

There is significant difference between the ZFC and the FC data in
the ordered state for both $H~\parallel~ab$ and $H~\parallel~c$,
with $\chi_c~>~\chi_{ab}$ in the low-temperature paramagnetic
state (inset Fig.\ref{F05}). This is consistent with the magnetic
moments ordering ferromagnetically along the \textit{c}-axis below
the irreversibility temperature $T_{irr}~=~63.8~\pm~1.9$ K, as it
has already been reported on polycrystalline samples by Watson
\textit{et al.}\cite{wat13}. The irreversibility temperature for
$H~\parallel~ab$ is slightly different ($\sim~65.2$ K). Specific
heat data is needed to determine the magnetic ordering
temperature, and using an on-set criterion (Fig.\ref{F06}), the
Curie temperature was determined to be $T_C~=~67.5~\pm~0.5$ K,
larger than the anisotropic $T_{irr}$ values. In turn, this value
is significantly lower than the previously reported ordering
temperature for the polycrystalline samples\cite{wat13}. We
believe that the discrepancies in the ordering temperature
estimates are due to the different criteria used for determining
$T_C$, as well as to the different types of samples used in the
measurements, with the single crystal data possibly being more
accurate.\\

\textbf{TbPtIn}\\

\begin{figure}
\begin{center}
\includegraphics[angle=0,width=120mm]{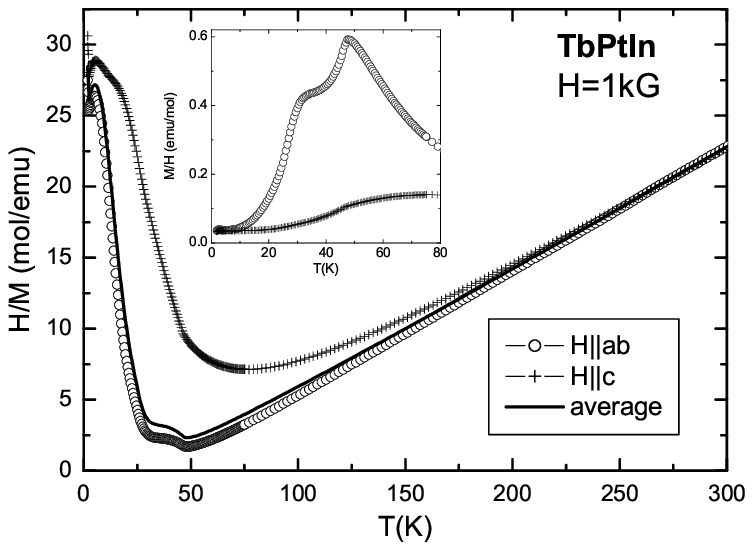}
\end{center}
\caption{Anisotropic $H~/~M$ data of TbPtIn and calculated average
(line) at H = 1 kG; inset: low-temperature anisotropic $M~/~H$
data.}\label{F08}
\end{figure}

We have already looked in detail at the magnetic and transport
properties of TbPtIn\cite{mor14}: in contrast to the neighboring R
= Gd member of the series, TbPtIn has an antiferromagnetic ground
state below $T_N~=~46.0$ K, with an extremely anisotropic, planar
susceptibility even in the paramagnetic state (Fig.\ref{F08}). At
higher temperatures, the inverse average susceptibility becomes
linear, indicating Curie-Weiss like behavior. Extrapolation of the
polycrystalline linear fit down to low temperatures yields an
effective moment value $\mu_{eff}~=~9.74~\mu_B$, in good agreement
with the theoretical value $9.72~\mu_{B}$ for Tb$^{3+}$ ions. The
anisotropic Weiss temperatures were also determined, and the
corresponding values are given in Table 1. Another phase
transition is apparent around $T_m~=~27.4$ K, possibly associated
with a spin reorientation. This phase transition was obscured in
the measurements on polycrystalline samples\cite{wat13}, whereas
the $T_N$ value that we determined based on measurements on single
crystals is fairly close to the previously reported one.

\begin{figure}
\begin{center}
\includegraphics[angle=0,width=120mm]{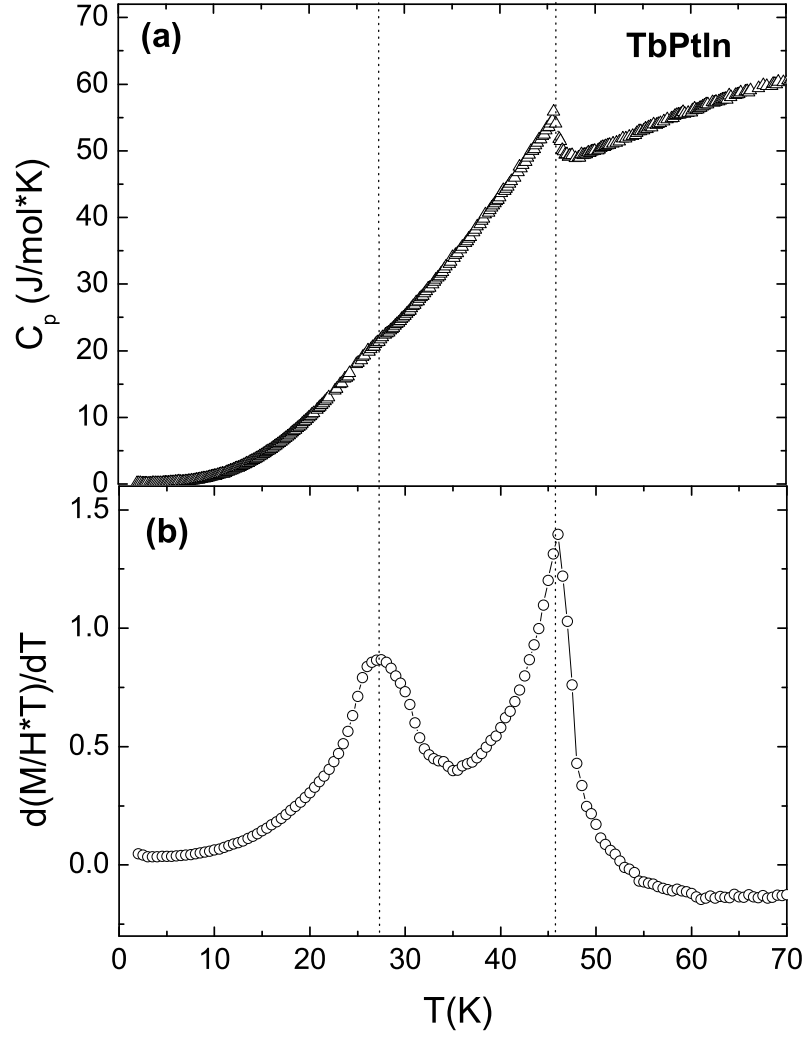}
\end{center}
\caption{(a) C$_P$(T) and (b) low-temperature $d(M_{ave}/H*T)/dT$
for TbPtIn; dotted lines mark the peak positions, corresponding to
the magnetic transitions.}\label{F09}
\end{figure}

As already seen by Morosan \textit{et al.}\cite{mor14}, the TbPtIn
specific heat shown in Fig.\ref{F09}a confirms the N\'{e}el
temperature and the lower temperature transition at $T_m$ (marked
by the vertical dotted lines). These transition temperatures are
also consistent with those revealed by the $d(M_{ave}/H*T)/dT$
data (Fig.\ref{F09}b), as expected for antiferromagnetic
compounds\cite{fis19}.

\begin{figure}
\begin{center}
\includegraphics[angle=0,width=120mm]{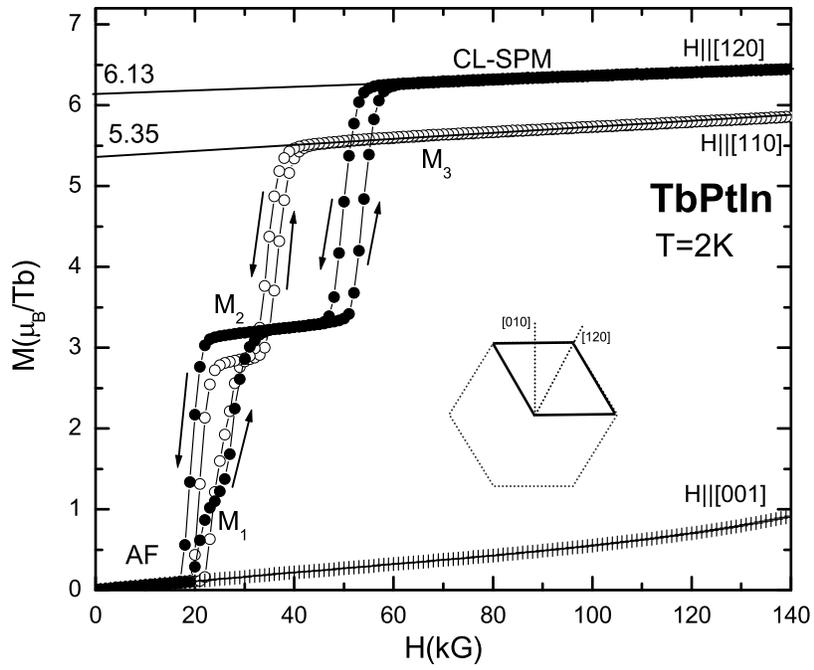}
\end{center}
\caption{Anisotropic field-dependent magnetization data for
TbPtIn, at T = 2 K, for increasing and decreasing field values (as
indicated by arrows).}\label{F10}
\end{figure}

Anisotropic field-dependent measurements at $T~=~2$ K
(Fig.\ref{F10}) reveal the presence of several metamagnetic
transitions for field applied perpendicular to the hexagonal
\textit{c}-axis, whereas for field along the \textit{c}-axis an
almost linear increase of the magnetization with field is observed
up to $\sim~140$ kG. As emphasized by Morosan \textit{et
al.}\cite{mor14}, apart from the extreme in-plane/out-of-plane
anisotropy, there is also a complex angular dependence of the
observed metamagnetism for $H~\perp~c$. The full and open circles
in Fig.\ref{F10} represent the measurements corresponding to the
two high symmetry in-plane directions of the applied field
(\textit{i.e.}, the [120] and [110] directions respectively), for
increasing and decreasing fields. The high-field magnetization
values reach $6.45~\mu_{B}~/$ Tb$^{3+}$ and $5.86~\mu_{B}$/
Tb$^{3+}$ for the two in-plane directions, and, within the
\textit{three coplanar Ising-like model}\cite{mor14}, correspond
to the crystal field limited saturated paramagnetic CL-SPM state.
Also consistent with this model in the low energy limit is the low
value of the axial component of the magnetization
$M([001])~=~0.92~\mu_{B}~/$ Tb$^{3+}$. However, a slow increase of
the high-field magnetization plateaus is apparent for $H~\perp~c$;
this, as well as the slow increase of $M_c$ with the applied
field, may be due to the fact that the system is approaching the
CEF splitting energy scale. Extrapolation of the high-field
magnetization data (solid lines in Fig.\ref{F10}) down to $H~=~0$
leads to smaller values (\textit{i.e.}, $6.13~\mu_{B}$,
$5.86~\mu_{B}$ and 0 for $M([120])$, $M([110])$, and $M([001])$
respectively), even closer to the theoretical ones\cite{mor14}.\\

\textbf{DyPtIn}\\

So far we have seen that GdPtIn has a ferromagnetic ground state,
with $\chi_c~>~\chi_{ab}$ in the low-temperature paramagnetic
state, whereas TbPtIn orders antiferromagnetically and is
extremely planar, even for a limited temperature range above
$T_N$.

\begin{figure}
\begin{center}
\includegraphics[angle=0,width=120mm]{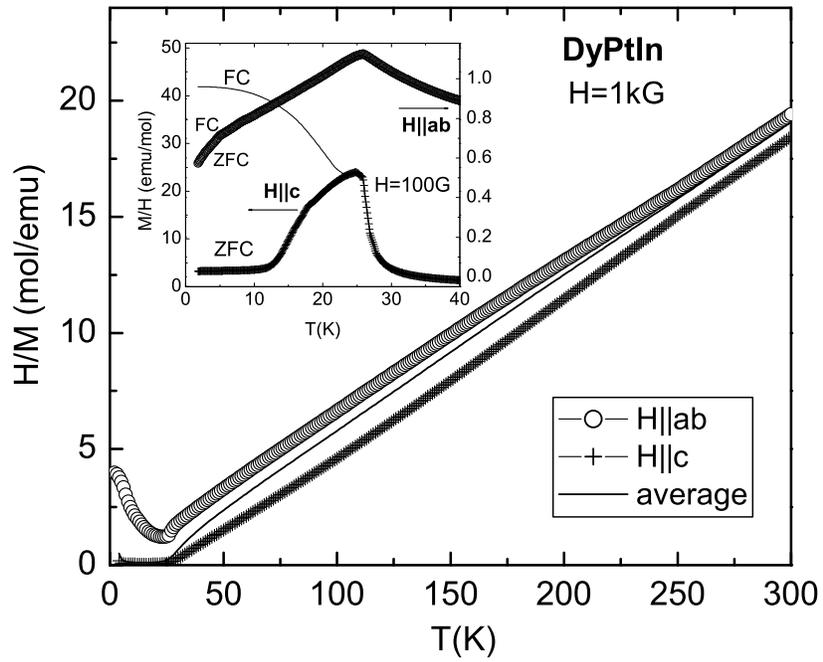}
\end{center}
\caption{Anisotropic $H~/~M$ data for DyPtIn and the calculated
average (line) at H = 1 kG; inset: ZFC-FC low-temperature
anisotropic $M~/~H$ data for H = 0.1 kG.}\label{F11}
\end{figure}

As we move towards the heavier R members of the series, DyPtIn
resembles more the R = Gd compound rather than the neighboring R =
Tb one: from the anisotropic $H~/~M$ data data shown in
Fig.\ref{F11}, it appears that DyPtIn has a linear inverse average
susceptibility, from which an effective moment
$\mu_{eff}~=~10.7~\mu_B$ can be determined, consistent with the
theoretical value of $10.6~\mu_B$. The anisotropic inverse
susceptibilities can also be used to determine the Weiss
temperatures, listed in Table.1 for both orientations of the
field, as well as for the polycrystalline average. Below $\sim~30$
K, DyPtIn orders magnetically (the Curie temperature T$_C$ will be
determined from the specific heat data, shown below). The
ordered-state $M~/~H$ data indicates a possible net ferromagnetic
component along the \textit{c}-axis. Moreover, ZFC and FC data for
$H~=~100$ G (inset, Fig.\ref{F11}) further confirm this
hypothesis, given the irreversibility of the $\chi_c$ data below
$\sim~25$ K, and no visible irreversibility for the $\chi_{ab}$
data.

\begin{figure}
\begin{center}
\includegraphics[angle=0,width=120mm]{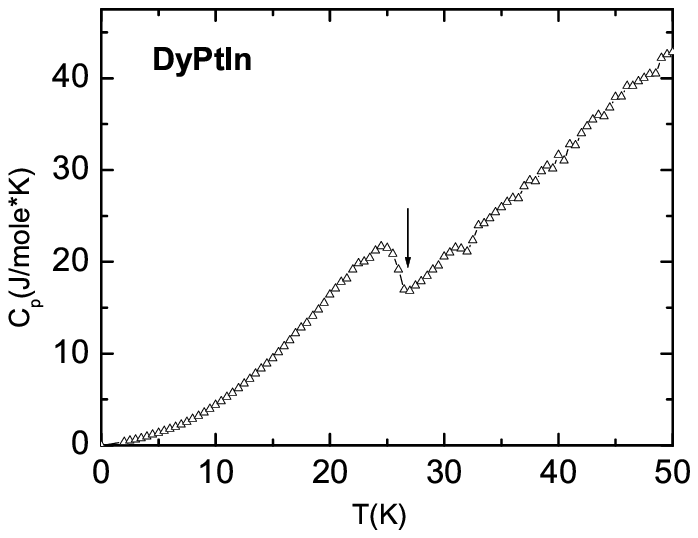}
\end{center}
\caption{Specific heat C$_P(T)$ of DyPtIn; small arrow indicates
T$_C$ determined from on-set (see text).}\label{F12}
\end{figure}

As previously seen for GdPtIn, a rather broad peak in the specific
heat data (Fig.\ref{F12}) indicates the magnetic ordering of the
DyPtIn. Using the on-set criterion, the Curie temperature is
determined to be $T_C~=~(26.5~\pm~0.5)$ K, indicated by the small
vertical arrow. The substantial difference between our estimates
and those of Watson et al.\cite{wat13} for the ordering
temperature, which appears to persist for all RPtIn members (R =
Gd - Dy) described so far, could be a consequence of the two sets
of data having been collected on single crystal or polycrystalline
samples, respectively. However, for the ferromagnetic compounds,
different criteria used for determining the ordering temperature
may also be causing the aforementioned differences.

\begin{figure}
\begin{center}
\includegraphics[angle=0,width=120mm]{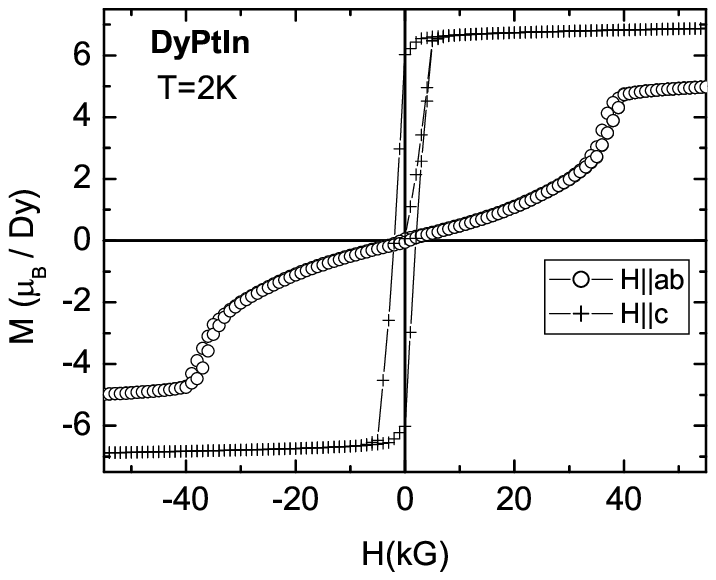}
\end{center}
\caption{Anisotropic field-dependent magnetization loops for
DyPtIn, at T = 2 K.}\label{F13}
\end{figure}

The field-dependent magnetization loops ($-~55$ kG
$\leq~H~\leq~55$ kG) are shown in Fig.\ref{F13}, for both
$H~\parallel~c$ and $H~\perp~c$. For the applied field along the
\textit{c}-axis (crosses), a small hysteresis can be observed,
whereas the magnetization rapidly increases towards a
saturated-like value around $\sim~6.88~\mu_B~/$ Dy$^{3+}$. This is
consistent with a ferromagnetic component of the magnetization
along the \textit{c}-axis, which is well below the expected
$10~\mu_B$ saturated value for Dy$^{3+}$ ions. For $H~\perp~c$, a
metamagnetic transition occurs around $\sim~37$ kG, leading to a
state with the magnetization value around 4.98 $\mu_B$, even
smaller than the axial component. As we shall see for the rest of
the local-moment members of the series (R = Ho, Er, Tm), the
measured values of the magnetization at the highest applied field
are far smaller than the theoretical saturated values for the
respective ions, for both $H~\parallel~c$ and $H~\perp~c$.
Starting from the two-dimensional model already developed for
TbPtIn\cite{mor14}, we will attempt to generalize it to three
dimensions such as to explain the nature of the ordered state
across the whole RPtIn series (R = Gd - Tm).\\

\textbf{HoPtIn}

HoPtIn has similar physical properties to GdPtIn and DyPtIn, and
appears to conform to some general characteristics of the RPtIn
series, with the Tb member as an apparent exception: axial
ferromagnetic component of the ordered state magnetization, less
than the theoretical saturated values above $\sim~50$ kG for both
the axial and the planar magnetizations.

\begin{figure}
\begin{center}
\includegraphics[angle=0,width=120mm]{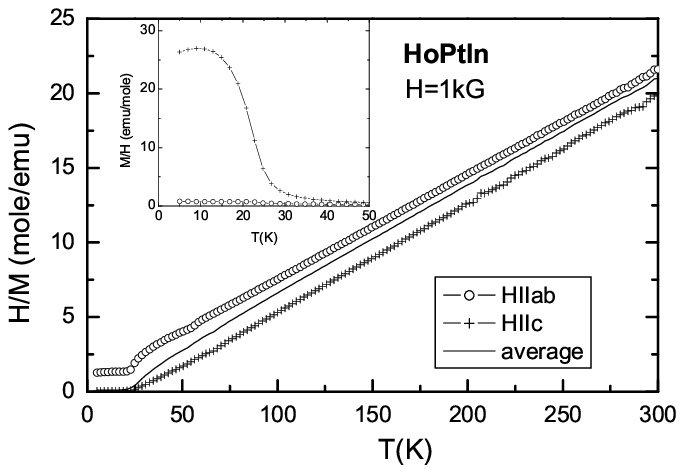}
\end{center}
\caption{Anisotropic $H~/~M$ data for HoPtIn and the calculated
average (line) at H = 1 kG; inset: low-temperature anisotropic
$M~/~H$ data.}\label{F14}
\end{figure}

\begin{figure}
\begin{center}
\includegraphics[angle=0,width=120mm]{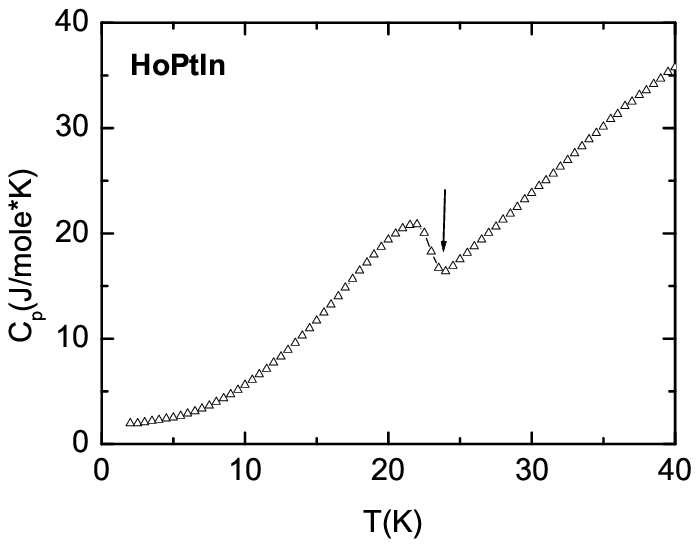}
\end{center}
\caption{Specific heat C$_P(T)$ of HoPtIn; small arrow indicates
T$_C$ determined from on-set (see text).}\label{F15}
\end{figure}

As can be seen in Fig.\ref{F14}, the anisotropic inverse
susceptibilities are linear at high temperatures; from the
polycrystalline average, we get an effective moment of
$10.5~\mu_B$, close to the theoretical value
$\mu_{eff}($Ho$^{3+})~=~10.6~\mu_B$. The presence of the
ferromagnetic component of the ordered state is evidenced by the
anisotropic $M~/~H$ data featuring a large, broad peak at low
temperatures for $H~\parallel~c$ (inset Fig.\ref{F14}), with
$\chi_c~>~\chi_{ab}$ in the low-temperature paramagnetic state. In
the specific heat data (Fig.\ref{F15}), magnetic ordering is
apparent below $T_C~=~23.5~\pm~0.5$ K, as indicated by the small
arrow.

The idea of a ferromagnetic component of the magnetization is
further confirmed by the field dependent data in Fig.\ref{F16},
where for the field applied in the \textit{c} direction (crosses),
the magnetization rapidly increases with H. The maximum value
reached within our field limits is $\sim~7.81~\mu_B$, less than
the calculated saturated moment for Ho$^{3+}$ ions. As the
magnetic field is applied parallel to the basal plane, the
resulting magnetization curve is consistent with either a broad
metamagnetic transition or with a continuous spin-flop transition.
Around $H~=~55$ kG, the in-plane component of the magnetization is
$4.3~\mu_B$, even smaller than the axial one and less than half of
$\mu_{sat}($Ho$^{3+})$.\\

\textbf{ErPtIn}\\

As TbPtIn appears to be an exception, the R = Er member of the
RPtIn series follows the already observed trends for the other
heavy R compounds. The $H~/~M$ average data seen in Fig.\ref{F17}
is linear towards high temperatures, indicative of Curie-Weiss
behavior of magnetization. However, crossing of the planar and
axial inverse susceptibilities occurs around 150 K, possibly a
result of strong crystal field effects at high temperatures in
this compound. Similar crossing of the anisotropic inverse
susceptibilities was also observed for the R = Er member of the
RNi$_2$B$_2$C series\cite{choEr}. The effective moment value
extracted from the high-T linear region of the inverse average
susceptibility is $\mu_{eff}~=~10.1~\mu_B$, close the theoretical
$9.6~\mu_B$ value (within the accuracy of our data and fit).

The low temperature anisotropic $H~/~M$ data shown in the inset is
consistent with ferromagnetic ground state, with
$\chi_c~>~\chi_{ab}$ in the low-T paramagnetic state. The Curie
temperature, as determined from the specific heat data in
Fig.\ref{F18}, is $T_C~=~8.5~\pm~0.5$ K, as the small arrow
indicates.

From the field-dependent  measurements in Fig.\ref{F19} we also
infer that the magnetization has a ferromagnetic component along
the \textit{c}-axis, as the corresponding data (crosses) rapidly
increase with field. Above $\sim~10$ kG, the axial magnetization
has an almost constant value around $7.50~\mu_B$, whereas the
theoretical saturated moment for Er$^{3+}$ ions is $9~\mu_B$. When
$H~\perp~c$ (open circles), the magnetization data is almost
linear in field, with a weak hint of un upward curvature around 20
kG, possibly indicating a metamagnetic transition. Towards 50 kG,
the magnetization almost levels off around a $2.77~\mu_B$ value,
much lower than the expected saturated moment.\\

\textbf{TmPtIn}\\

Having an antiferromagnetic ground state and a planar
magnetization component larger than the axial one, TbPtIn differs
from the rest of the RPtIn compounds mentioned so far, whereas
below we show that it resembles the R = Tm member of this series.

The high temperature inverse susceptibility of TmPtIn
(Fig.\ref{F20}) is linear, yielding an effective magnetic moment
around $7.7~\mu_B$, close to the theoretical value calculated for
Tm$^{3+}$ ions, $\mu_{eff}~=~7.6~\mu_B$. However, unlike the
aforementioned members of the series (except Tb), below $\sim~4$ K
this compound appears to order antiferromagnetically, as suggested
by the low-temperature susceptibility data in the inset. Sharp
peaks in the susceptibility data around $T_N$ are typically
indicative of antiferromagnetic ordered state, as is the case with
the $H~\parallel~c$ data (crosses) shown in the inset in
Fig.\ref{F20}. The similar peak for the in-plane susceptibility
(open circles) is somewhat broader, possibly due to spin
fluctuations or CEF effects, which result in reduced
susceptibility values around the ordering temperature.

A peak in the specific heat (Fig.\ref{F21}a) suggests that the
magnetic order occurs at $T_N~=~3.0~\pm~0.5$ K, and, as expected
for antiferromagnetic compounds\cite{fis19}, is consistent with
the $d(M_{ave}/H*T)/dT$ data in Fig.\ref{F21}b.

The $T~=~2$ K magnetization isotherms (Fig.\ref{F22}) indicate one
(for $H~\parallel~c$) or more (for $H~\parallel~ab$) metamagnetic
transitions. Following these fairly broad transitions (due to the
high temperature, compared to $T_N$, for which these data were
taken), the magnetization curves seem to approach some horizontal
plateaus around $2.26~\mu_B$ for the field along the
\textit{c}-axis, and $4.42~\mu_B$ for the field within the
\textit{ab}-plane respectively. As for the other RPtIn (except for
R = Gd), both these values are much smaller than the calculated
effective moment of $7~\mu_B$ for the Tm$^{3+}$ ions.\\

\subsection{YbPtIn}

YbPtIn stands out from the rest of the RPtIn compounds through a
number of distinctly different properties. Fig.\ref{F23} shows the
inverse anisotropic susceptibilities (symbols), together with the
calculated polycrystalline average (solid line). The latter data
is linear above $\sim~50$ K, despite a pronounced departure from
linearity of the axial inverse $H~/~M$ data (crosses) below
$\sim~200$ K, probably due to CEF effects. From the fit of the
linear part of the average $H~/~M$ data, an effective moment of
4.3 $\mu_B/$ Yb$^{3+}$ can be determined, close to the theoretical
4.5 $\mu_B/$ value. For lower temperatures, no distinguishable
features associated with magnetic order are visible in the $M~/~H$
data data down to 1.8 K (inset, Fig.\ref{F23}), for field values
of 0.1 and 20 kG. This observations are consistent with the
susceptibility data reported by Kaczorowski \textit{et
al.}\cite{kac10}. However, the specific heat data (Fig.\ref{F24})
shows a sharp peak around 2.1 K, and the feature associated with
this transition could have been missed in the $M~/~H$ measurements
because of the limited temperature range below the transition.

Trovarelli \textit{et al.}\cite{tro09} presented magnetization
measurements that, for low applied fields ($H~=~0.1$ kG), suggest
antiferromagnetic ordering below $T_N~=~3.4$ K, a value that is
different from the possible transition temperature indicated by
our measurements. In trying to understand the possible cause of
such differences, single crystal x-ray measurements were
performed. They indicate that our flux-grown YbPtIn single
crystals have a partial (\textit{i.e.}, $\sim~94~\%$) occupancy
for one of the two Pt sites in the unit cell, such that the
stoichiometry of these crystals is closest to YbPt$_{0.97}$In.
This is not entirely surprising, given the different flux growth
process (a low Pt-concentration used in the initial
Yb$_{0.4}$Pt$_{0.1}$In$_{0.5}$ solution). Consequently the R = Yb
compound is excluded from the following discussion. A more
complete analysis of the thermodynamic and transport properties of
this material is currently the subject of a different
study\cite{mor21}.

\section{data analysis}

The magnetic RPtIn compounds that we investigate here appear to
order magnetically below $\sim~70$ K. As can be seen in
Fig.\ref{F25}, their ordering temperatures $T_{ord}$ scale fairly
well with the de Gennes factor $dG~=~(g_J~-~1)^2~J~(J~+~1)$, where
$g_J$ is the Land\'{e} g-factor, and J is the total angular
momentum of the R$^{3+}$ ion Hund's rule ground state. Whereas
this suggests that the RKKY interaction between the conduction
electrons and the local magnetic moments gives rise to the
long-range magnetic order, slight departures from linearity,
similar to those seen for other rare
earth-series\cite{mye21,bud22,mor20}, are due to the extremely
simplified assumptions associated with the de Gennes scaling. The
scaling is apparently unaffected by the curious switching from
ferromagnetic to antiferromagnetic ordering across the RPtIn
series, which appears to be correlated with a change in the
anisotropy, such that, in the low-T paramagnetic state,
$\chi_c~>~\chi_{ab}$ for the ferromagnetic compounds, and
$\chi_{ab}~>~\chi_c$ for the antiferromagnetic ones. At first,
this may seem inconsistent with de Gennes scaling, which would
indicate similar ordering mechanisms for all RPtIn compounds, R =
Gd - Tm. As we shall see, we believe that, because of their
Fe$_2$P-type hexagonal structure, with three R ions sitting at
orthorhombic point symmetry sites, strong CEF effects constrain
the local magnetic moments in R = Tb - Tm to equivalent
non-collinear easy-axes. This results in (i) anisotropic
paramagnetic magnetization, and (ii) crystal field limited
saturated paramagnetic CL-SPM states with magnetization values
well below the corresponding free ion saturated moments.

We have already modelled the effects of strong crystal electric
fields on the Fe$_2$P-type crystal structure, for the case of the
extremely planar R = Tb member of the RPtIn series and the similar
R = Tm member of the RAgGe series, using the \textit{three
co-planar Ising-like systems model}\cite{mor14}: having three rare
earths in orthorhombic point symmetry, the hexagonal symmetry of
the unit cell was achieved by constraining the local moments to
three equivalent co-planar directions, $60^0$ away from each
other. In allowing both the "up" and "down" positions
(\textit{i.e.}, Ising-like) for each of the three magnetic
moments, the antiferromagnetic ground state was, in the simplest
case, realized by a ($\searrow\uparrow\swarrow$) moment
configuration (Fig.\ref{F26}a); upon increasing the applied
magnetic field within the basal plane, a number of metamagnetic
states occurred, showing simple dependencies of the critical
fields $H_c$ and the locally saturated magnetizations $M_{sat}$ on
the angle $\theta$ between the direction of the field and the easy
axis (see Fig.13 and related discussion by  Morosan \textit{et
al.}\cite{mor14}). When all the moments are in their "up"
positions ($\nwarrow\uparrow\nearrow$), a crystal field limited
saturated paramagnetic CL-SPM state is reached (Fig.\ref{F26}b);
the expected magnetization value is

$1/3*\mu_{sat}($Tb$^{3+})*(1+2*\cos60^0)~=~2/3*\mu_{sat}($Tb$^{3+})$
or $2/3*9~\mu_B~=~6~\mu_B$,

consistent with the easy-axis measured data (full circles) shown
in Fig.\ref{F10}.

We thus see that the aforementioned \textit{three co-planar
Ising-like systems model} explains how the measured magnetization
values can be much smaller than the theoretical saturated value of
$9~\mu_B$ for the Tb$^{3+}$ ions.

By contrast, the GdPtIn doesn't exhibit such reduced values of the
magnetization for high applied fields (Fig.\ref{F07}), given that
the Gd$^{3+}$ ions are in a symmetric $^8S_{7/2}$ state, and thus
the CEF effects are minimal: for $H~\parallel~c$ (crosses), the
magnetization rapidly increases, reaching
$\mu_{sat}($Gd$^{3+})~=~7~\mu_B$ for $H~\geq~10$ kG. This is
typical of a ferromagnetic magnetization for field applied along
the direction of the moments (easy axis). Furthermore, the
$H~\perp~c$ data (open circles) represent classical hard axis
data, and are consistent with axial ferromagnetic ordering in this
compound, as the saturated state is also reached, however at a
slower rate (\textit{i.e.}, for $H~\geq~40$ kG).

For the other neighboring TbPtIn compound, the DyPtIn
magnetization resembles the similar data for GdPtIn, even though
the presence of CEF effects in the former system results in
reduced magnetization values at our maximum applied field: as can
be seen in Fig.\ref{F13}, the $H~\parallel~c$ magnetization
(crosses) rapidly increases with field, as expected for a
ferromagnet with moments along \textit{c}, but at $H~=~55$ kG its
value is only $\sim~0.7$ of the theoretical saturated moment of 10
$\mu_B$. For field applied within the basal plane (open circles),
only 0.5 of the saturated moment is reached following a
metamagnetic transition around 35 kG. Whereas more metamagnetic
transitions beyond our maximum field of 55 kG could account for
the small magnetization values in this compound, such a hypothesis
does not address one more peculiarity already apparent for the
RPtIn series: even though the R = Gd and Dy compounds are
ferromagnetic, and the R = Tb one is antiferromagnetic, their
ordering temperatures scale well with the de Gennes factor, as we
showed in Fig.\ref{F25}. Furthermore, the R = Ho and Er compounds
also display ferromagnetic components of the ordered state
magnetization, whereas TmPtIn is antiferromagnetic, and yet the de
Gennes scaling still holds for all heavy RPtIn compounds
(Fig.\ref{F25}). The question arises whether a generalized
hypothesis exists, which could account for the magnetic ordering
in all RPtIn systems (R = Gd - Tm), or whether TbPtIn and TmPtIn
should be regarded as exceptions from the ferromagnetic axial
ordering across the series.

In what follows we will present one plausible model for the
magnetic ordering in the local moment RPtIn compounds, a
generalized version of the two-dimensional \textit{three
Ising-like systems model}, which addresses the above points. We
first proposed such a model for DyAgGe\cite{mor20}, an
isostructural compound to RPtIn, for which a ferromagnetic
component of the magnetization was also apparent along the
\textit{c}-axis.

In the \textit{three co-planar Ising-like systems model}, we
assume that the magnetic moments are allowed to three orientations
(along any three of the six equivalent six-fold symmetry axes
within the basal plane), with two positions ("up" and "down") per
orientation. This results in a set of three two-fold degenerate
easy-axes, $60^0$ away from each other, the degeneracy being a
direct consequence of the requirement that the Ising-like systems
be co-planar: the "up" position for a given easy-axis is
indistinguishable from the "down" position for the equivalent
direction $3*60^0~=~180^0$ away.

If we release the restriction that the moments be co-planar, while
still imposing that their \textit{in-plane} projections be $60^0$
away from each other to preserve the hexagonal symmetry of the
crystals, this degeneracy is lifted, and the moments are not
necessarily Ising-like systems any more. The three-dimensional
model described above can be directly derived from the planar one,
as follows: we consider that Fig.\ref{F26} represents the
\textit{in-plane} projection of the magnetic moments'
configuration, to which non-zero axial components of the moments
are added. The possible resulting moment configurations can be
obtained using any combination of "up" (thin solid arrows) or
"down" (thin dotted arrows) planar and axial components of the
magnetic moments, as shown in Fig.\ref{F27}. Given the
orthorhombic point symmetry of the magnetic moments' sites, this
yields two possible co-planar orientations for each magnetic
moment, with the corresponding "up" and "down" positions for each.
The thick solid arrows in Fig.\ref{F27} represent the full
magnetic moments, which are parallel to three non-planar,
equivalent directions (\textit{i.e.}, easy axes), inclined at an
angle $\alpha$ from the \textit{c}-axis. This configuration
corresponds to the crystal field limited saturated paramagnetic
CL-SPM state, where all \textit{in-plane} and axial components of
the magnetic moments are in their respective "up" positions.

It is worth noting that, by analogy with the two-dimensional
model, there are two sets of easy axes: the
[1~2~\textit{l}]-equivalent axes, where \textit{l} is the
\textit{c}-axis Miller index, for which the corresponding
\textit{in-plane} model exactly describes TbPtIn, or the
[1~1~\textit{l}]-equivalent directions, with a two-dimensional
analogous example being TmAgGe (also described in detail by
Morosan \textit{et al.}\cite{mor14}). In our present model and
data analysis, we are assuming the first scenario, in which the
easy axes are the [1~2~\textit{l}]-equivalent directions, which
project in the \textit{ab}-plane onto the [1~2~0] directions.
Consequently we will refer to the (1~2~0) planes as the "easy"
planes and the [1~2~0] directions as "easy" \textit{in-plane}
axes, whereas the (1~1~0) planes and the [1~1~0]-axes will be
called in this case "hard" planes and "hard" \textit{in-plane}
axes respectively. When the [1~1~\textit{l}] directions are the
easy axes, the same description is still valid, with the "easy"
and "hard" planes and \textit{in-plane} directions interchanged
from the previous case.

We have thus introduced a model generalized from the \textit{three
co-planar Ising-like systems model}, which takes into account the
CEF effects on hexagonal compounds with orthorhombic point
symmetry of the rare earths. For each compound, the strength of
the CEF effects will be reflected by the value of the angle
$\alpha$ between the easy-axes and the \textit{c}-axis. At low
temperatures, another energy scale is introduced by the applied
magnetic field, and the model described above is only valid for
fields much smaller than the CEF energy. In this limit, for the
highest applied fields, a CL-SPM state is reached, for which the
anisotropic magnetization values are smaller than the theoretical
saturated moments $\mu_{sat}$ for the respective R ions.

For a fixed angle $\alpha$, there are six possible easy-axes (or
three pairs of co-planar easy axes), each with the corresponding
"up" and "down" positions. As in the case of the two-dimensional
model, multiples S of three moments may be required to
characterize the moment configuration for a given applied field.
The orientation of the applied field will determine the magnetic
moments to align along the three easy axes closest to the
direction of the field, whereas its magnitude will determine the
number of "up" and "down" moments along those three easy axes.

The CL-SPM state is reached when all three magnetic moments are in
their "up" positions along three adjacent easy axes closest to the
field direction (or, equivalently, when all \textit{in-plane} and
axial components of the magnetic moments are in their "up"
positions). This state is illustrated in Fig.\ref{F27}, for the
magnetic field applied off the \textit{c}-axis. (If H is parallel
to the \textit{c}-axis, all six "up" positions of the magnetic
moments are equally probable, and only when rotating the field
away from \textit{c} the three easy axes closest to the applied
field direction are uniquely determined).

Experimentally we can only measure the projection of the magnetic
moments along the field direction, with the resulting
magnetization per magnetic moment given by

$M~=~\frac{1}{3}~[~\overrightarrow{M}_1~+~\overrightarrow{M}_2~+~\overrightarrow{M}_3~]~\cdot~\frac{\overrightarrow{H}}{H}$.

Moreover, we were able to measure the angular dependent
magnetization for the magnetic field applied within the horizontal
\textit{ab}-plane and the high-symmetry vertical planes
(\textit{i.e.}, "easy" or "hard" planes). Such data can be used to
probe the validity of our model, by comparison with the
theoretical calculation of the expected angular dependent "easy"
and "hard" magnetization values.

For a fixed angle $\alpha$, and for field making an angle $\theta$
with the \textit{c}-axis, the magnetization values $M^e(\theta)$
and $M^h(\theta)$ in the CL-SPM state are, as calculated in detail
in the Appendix:

$M_{CL-SPM}^e/\mu_{sat}($R$^{3+})~=$

$~~~~~~~~~~~~~~~~~~~~~~~~~~~\frac{2}{3}*\sin\alpha*~\sin\theta~+~\cos\alpha*~\cos\theta$

and

$M_{CL-SPM}^h/\mu_{sat}($R$^{3+})~=$

$~~~~~~~~~~~~~~~~~~~~~~~~~~~\frac{\sqrt{3}}{3}*\sin\alpha*~\sin\theta~+~\cos\alpha*~\cos\theta$,

where $\theta$ is the angle between the applied magnetic field and
the \textit{c}-axis, and the indexes "e" and "h" denote,
respectively, the "easy"- and "hard"-plane components of the
magnetization. As already mentioned, we assume the "easy" and
"hard" axes to be the [1 2 0] and the [1 1 0] directions,
respectively. In what follows, our analysis refers only to the
CL-SPM state, therefore the subscript denoting the respective
state has been dropped for clarity.

From these calculations, the expected magnetization components (in
units of $\mu_{sat}($R$^{3+})$) for field parallel or
perpendicular to the \textit{c}-axis are:

$M([001])=M^e(\theta=0^0)=M^h(\theta=0^0)=\cos\alpha~\leq~1$,

$M([120])~=~M^e(\theta~=~90^0)~=~\frac{2}{3}~\sin\alpha~<~1$

and

$M([110])~=~M^h(\theta~=~90^0)~=~\frac{\sqrt{3}}{3}~\sin\alpha~<~1$.

Moreover, local maxima for the $M^e$ and $M^h$ curves are reached
for $\theta_{max}~=~\arctan(\frac{2}{3}~\tan\alpha)$, and
$\arctan(\frac{\sqrt{3}}{3}~\tan\alpha)$ respectively, with the
corresponding magnetization values equal to
$\sqrt{\cos^2\alpha~+~(\frac{2}{3})^2~\sin^2\alpha}~=~\sqrt{1-\frac{5}{9}~\sin^2\alpha}~<~1$,
and
$\sqrt{\cos^2\alpha~+~(\frac{\sqrt{3}}{3})^2~\sin^2\alpha}~=~\sqrt{1-\frac{6}{9}~\sin^2\alpha}~<~1$.

As can be seen from the above calculations, one should expect the
measured magnetization values to be smaller than the theoretical
saturated moment $\mu_{sat}($R$^{3+})$, regardless of the
direction of the applied field. The only exception is the axial
magnetization $M([001])$ for $\alpha~=~0^0$ (moments parallel to
the \textit{c}-axis), when the expected value is exactly
$\mu_{sat}($R$^{3+})$. These observations lend support to the idea
that the three-dimensional model considered above could describe
the RPtIn compounds, since for all R = Tb - Tm we have indeed
observed reduced values of the high-field anisotropic
magnetizations. On the other hand, it appears that the fully
saturated magnetization measured for GdPtIn could be described by
the above model for $\alpha~=~0^0$, but the absence of CEF effects
restricts the applicability of our model to this compound.

In the case of TbPtIn the magnetization measurements revealed
extreme planar anisotropy of this compound (Fig.\ref{F08} and
\ref{F10}). Within our three-dimensional model, this is consistent
with the angle $\alpha$ being equal to $90^0$, when the magnetic
moments become co-planar. In this case, the calculated
magnetization values become (in units of $\mu_{sat}($Tb$^{3+})$)

$M([001])=~\cos90^0~=~0$,

$M([120])~=~\frac{2}{3}~\sin90^0~=~\frac{2}{3}$

and

$M([110])~=~\frac{\sqrt{3}}{3}~\sin90^0~=~\frac{\sqrt{3}}{3}$.

These values show that when $\alpha~=90^0$, our model indeed
reduces to the \textit{three co-planar Ising-like systems
model}\cite{mor14}.

When $0^0~\leq~\alpha~<~90^0$, our model yields axial
magnetization values larger than 0, and this is observed for all
RPtIn, R = Tb - Tm. Therefore we can verify the applicability of
our model to these systems by estimating the angle $\alpha$, and
comparing the measured and calculated magnetization values in the
CL-SPM state as follows:

For all RPtIn compounds, the in-plane magnetization measurements
were performed for field along the [1 2 0] direction. Since it is
not readily apparent whether this represents the "easy" or "hard"
in-plane direction, one way to estimate the angle $\alpha$ is from
the $M([0~0~1])$ data:

$M([0~0~1])~/~\mu_{sat}($R$^{3+)}~=~\cos\alpha$.

Therefore $\alpha~=~\arccos(M([0~0~1])~/\mu_{sat}($R$^{3+}))$.
These values are listed in Table 2, together with the measured
magnetization values M([0~0~1] at $H~=~55$ kG, which were used in
the above formula.  For this value of the angle $\alpha$, the
"easy" and "hard" in-plane magnetization components should be, as
described above,

$M^e~/~\mu_{sat}($R$^{3+})~=~\frac{2}{3}~\sin\alpha$

and

$M^h~/~\mu_{sat}($R$^{3+})~=~\frac{\sqrt{3}}{3}~\sin\alpha$.

However, slight misalignments of the samples can occur for
$H~\parallel~c$, which may result in significant errors in our
determination of angle $\alpha$. In order to minimize these
errors, another way to determine $\alpha$ is from the ratio of the
two anisotropic measured magnetizations:

$M([1~2~0])~/~M([0~0~1])~=~2/3~*~\sin\alpha~/~\cos\alpha$ if the
[1 2 0] direction is the easy axis,

or
$M([1~2~0])~/~M([0~0~1])~=~\sqrt{3}/3~*~\sin\alpha~/~\cos\alpha$
if the [1 2 0] direction is the hard axis. The measured M([1~2~0])
values which were used for these calculations are listed in Table
2.

Thus the angle $\alpha$ is either

$\arctan~(~3~/~2~*~M([1~2~0])~/~M([0~0~1]))$

or $\arctan~(~3~/~\sqrt{3}~*~M([1~2~0])~/~M([0~0~1]))$,

and these estimated values are also listed in Table 2 as
$\alpha^e$ and $\alpha^h$.

As can be observed from the angle values listed in \\Table 2,
together with the error bars resulting from the two different
calculations, the angle $\alpha$ ranges from $89^0$ for TbPtIn, to
$\sim~32^0$ for ErPtIn. Our three-dimensional model seems to be
consistent with the experimental data for all RPtIn, R = Tb - Tm.

\begin{figure}
\begin{center}
\includegraphics[angle=0,width=120mm]{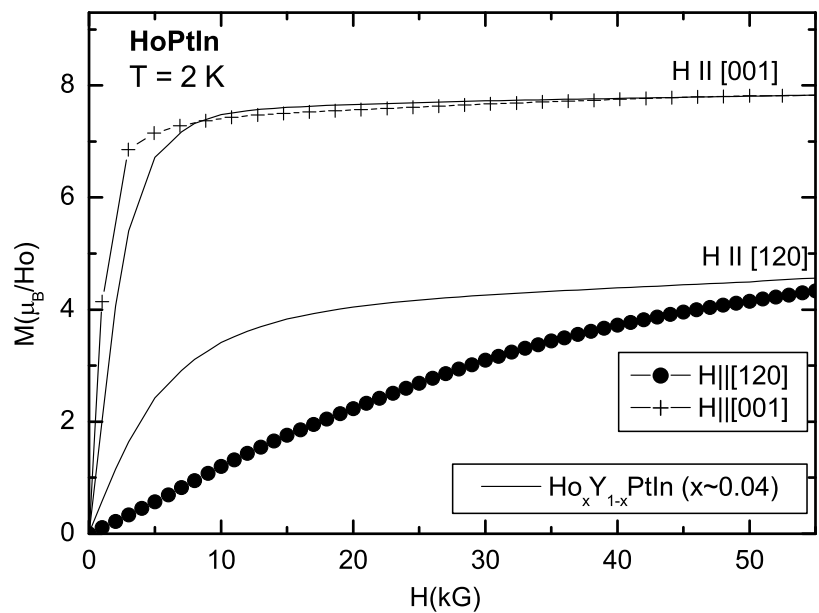}
\end{center}
\caption{Anisotropic field-dependent magnetization curves for
HoPtIn (symbols), and Ho$_x$Y$_{1-x}$PtIn, x $\sim~0.04$ (lines),
at T = 2 K.}\label{F16}
\end{figure}

\begin{figure}
\begin{center}
\includegraphics[angle=0,width=120mm]{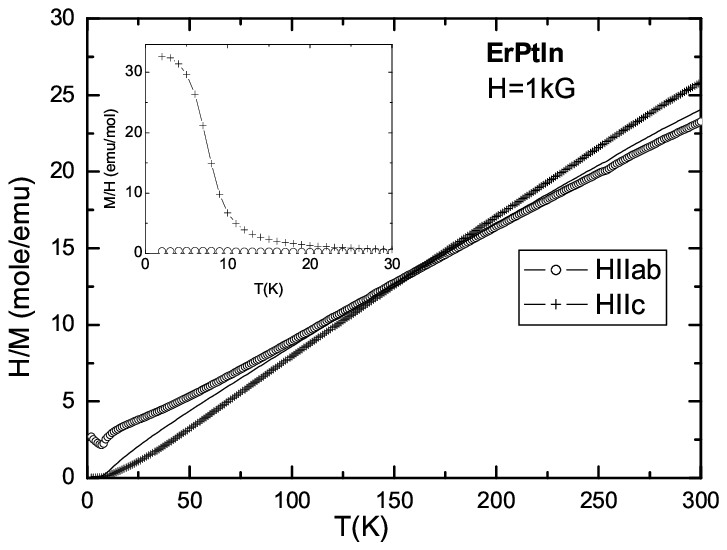}
\end{center}
\caption{Anisotropic $H~/~M$ data for ErPtIn and the calculated
average (line) at H = 1 kG; inset: low-temperature anisotropic
$M~/~H$ data.}\label{F17}
\end{figure}

\begin{figure}
\begin{center}
\includegraphics[angle=0,width=120mm]{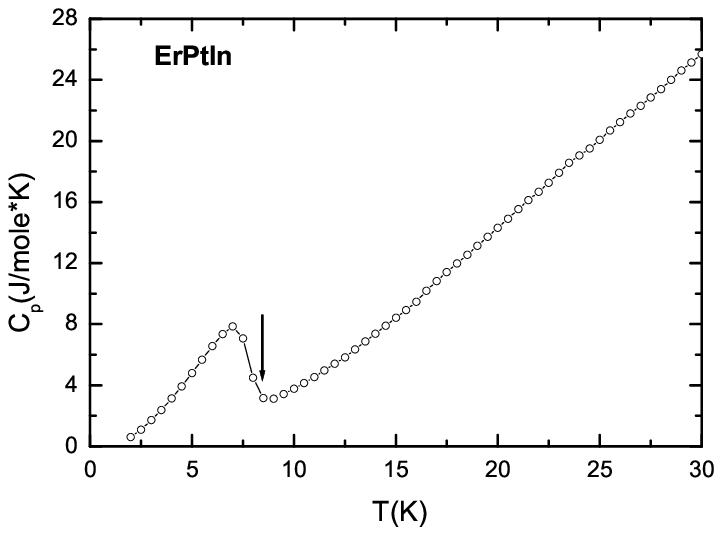}
\end{center}
\caption{Specific heat C$_P(T)$ of ErPtIn; small arrow indicates
T$_C$ determined from on-set (see text).}\label{F18}
\end{figure}

\begin{figure}
\begin{center}
\includegraphics[angle=0,width=120mm]{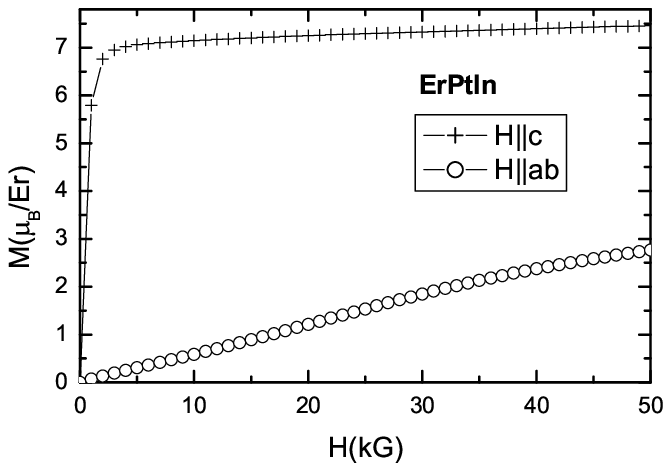}
\end{center}
\caption{Anisotropic field-dependent magnetization curves for
ErPtIn, at T = 2 K.}\label{F19}
\end{figure}

\begin{figure}
\begin{center}
\includegraphics[angle=0,width=120mm]{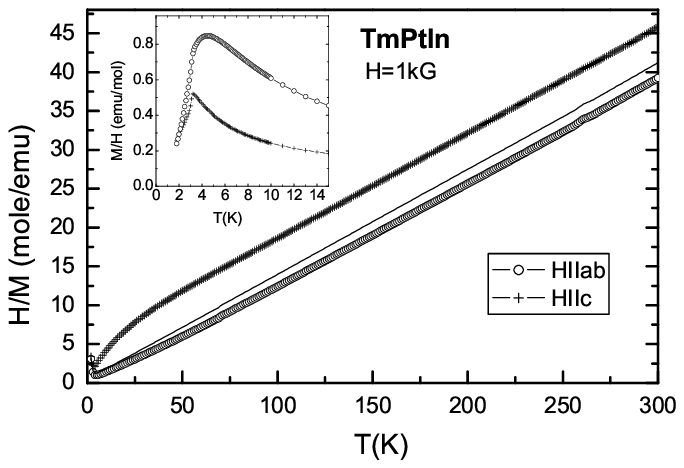}
\end{center}
\caption{Anisotropic $H~/~M$ data for TmPtIn and the calculated
average (line) at H = 1 kG; inset: low-temperature anisotropic
$M~/~H$ data.}\label{F20}
\end{figure}

\begin{figure}
\begin{center}
\includegraphics[angle=0,width=120mm]{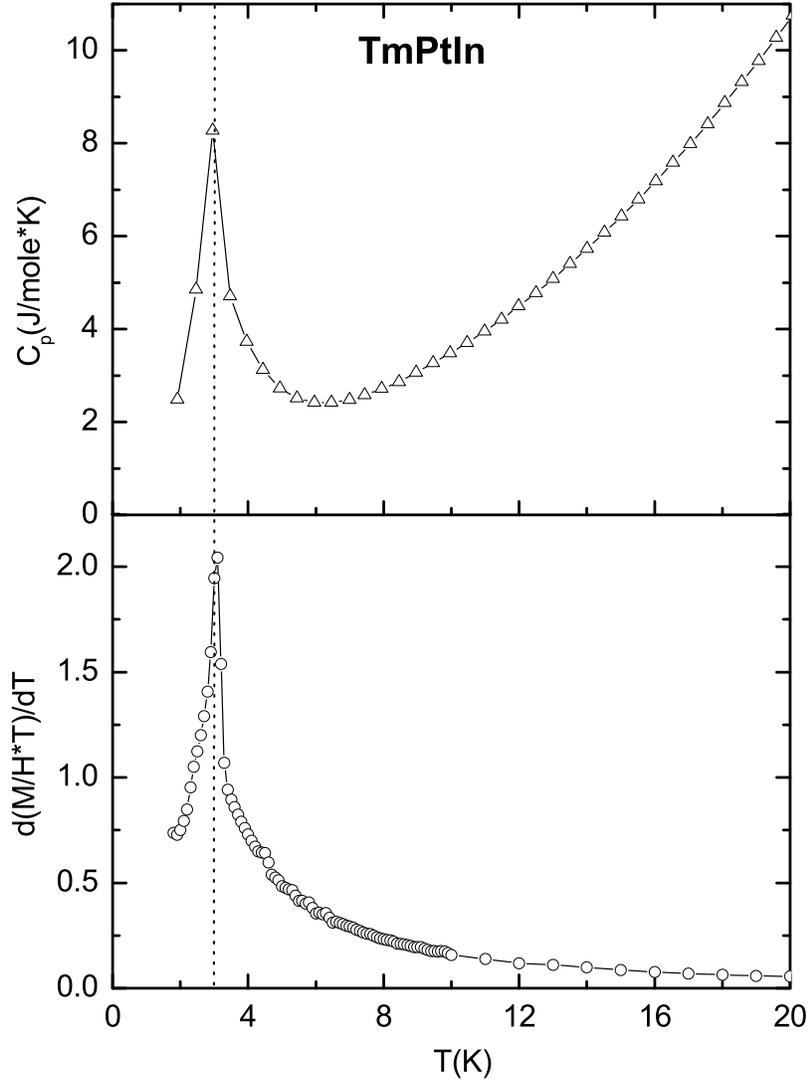}
\end{center}
\caption{(a) C$_P$(T) and (b) low-temperature $d(M_{ave}/H*T)/dT$
for TmPtIn; dotted line marks the peak position, corresponding to
T$_N$.}\label{F21}
\end{figure}

\begin{figure}
\begin{center}
\includegraphics[angle=0,width=120mm]{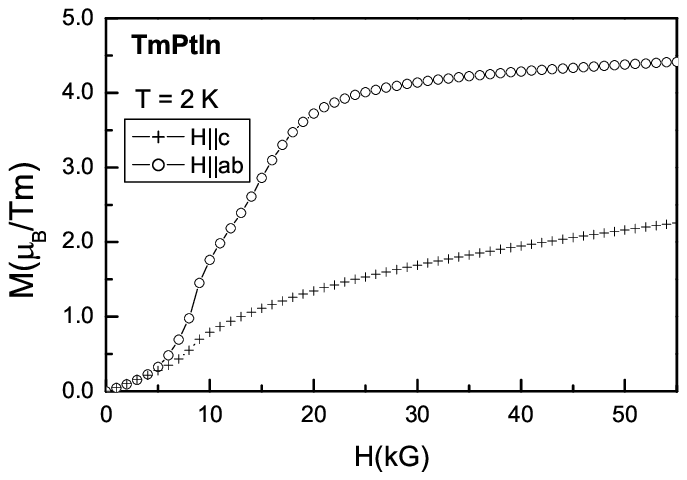}
\end{center}
\caption{Anisotropic field-dependent magnetization curves for
TmPtIn, at T = 2 K.}\label{F22}
\end{figure}

\begin{figure}
\begin{center}
\includegraphics[angle=0,width=120mm]{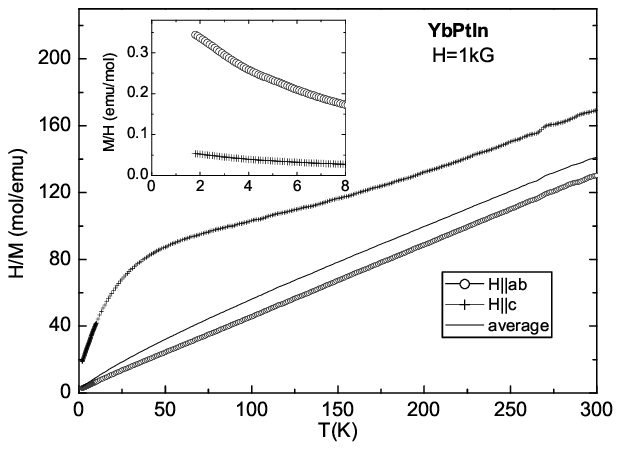}
\end{center}
\caption{Anisotropic $H~/~M$ data for YbPtIn and the calculated
average (line) at H = 1 kG; inset: low-temperature anisotropic
$M~/~H$ data.}\label{F23}
\end{figure}

\begin{figure}
\begin{center}
\includegraphics[angle=0,width=120mm]{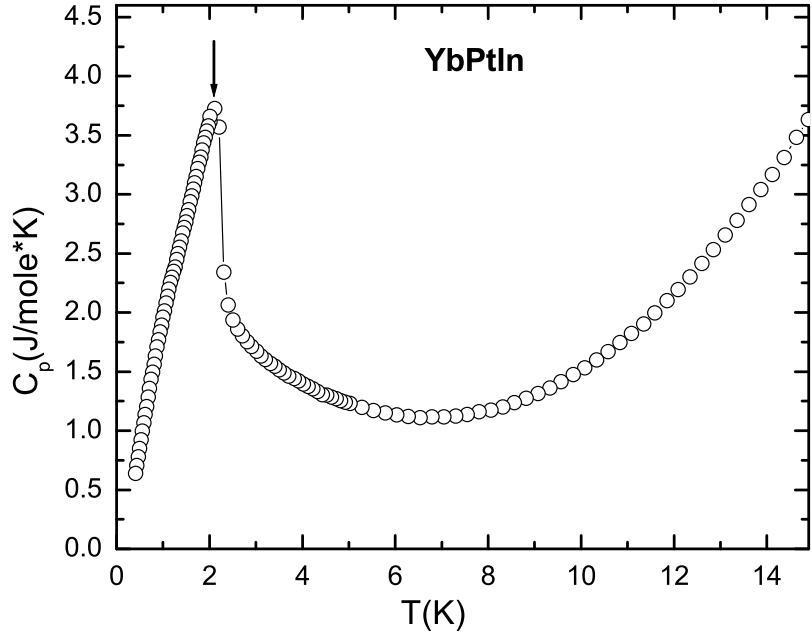}
\end{center}
\caption{Specific heat C$_P(T)$ of YbPtIn; small arrow indicates
T$_m$.}\label{F24}
\end{figure}

\begin{figure}
\begin{center}
\includegraphics[angle=0,width=120mm]{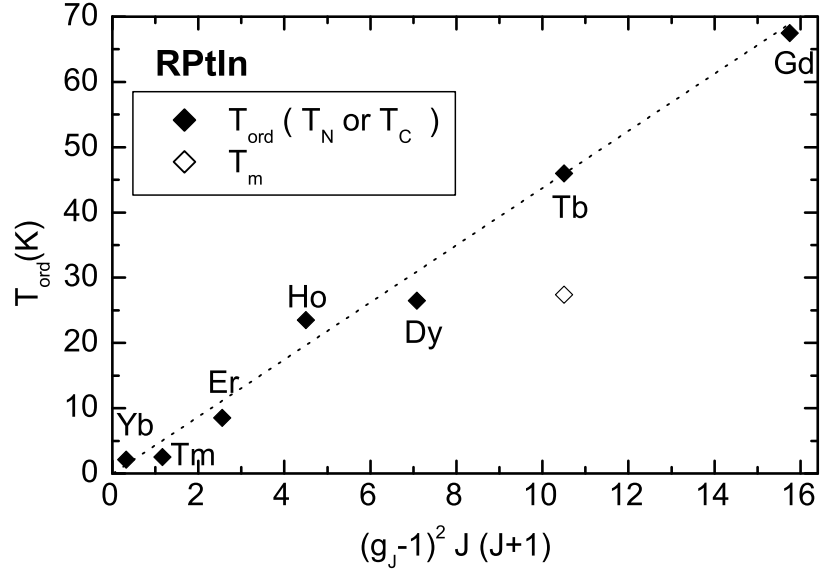}
\end{center}
\caption{Changes of the magnetic ordering temperatures T$_m$ for
RPtIn (R = Gd - Tm) with the deGennes scaling factor dG (the
dotted line represents the expected linear dependence). Open
symbol (for R = Tb) represents the low temperature transition from
the higher-T to lower-T antiferromagnetic state.}\label{F25}
\end{figure}

\begin{figure}
\begin{center}
\includegraphics[angle=0,width=120mm]{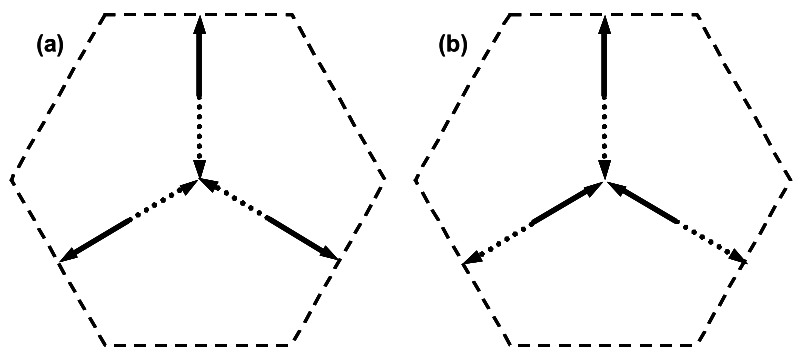}
\end{center}
\caption{Schematic representation of the three co-planar
Ising-like systems model in (a) the antiferromagnetic and (b) the
CL-SPM state. Solid arrows: "up" and dotted arrows: "down"
orientations of the magnetic moments along the easy
axes.}\label{F26}
\end{figure}

\begin{figure}
\begin{center}
\includegraphics[angle=0,width=120mm]{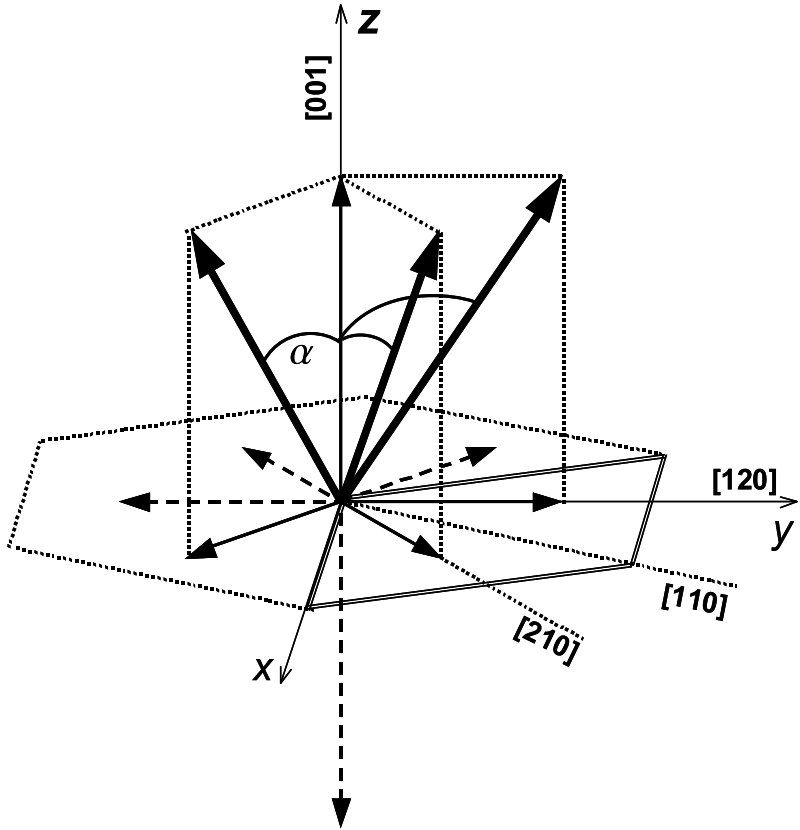}
\end{center}
\caption{The three-dimensional model for three magnetic moments at
unique orthorhombic point symmetry sites: thin arrows (solid -
"up" and dotted - "down") represent the non-zero components of the
magnetic moments along the [001] or the easy in-plane directions
(as shown, the [120] - equivalent directions). Thick arrows: full
magnetic moments in the CL-SPM state.}\label{F27}
\end{figure}

In order to further explore the validity of the above model,
angular dependent magnetization measurements were performed for a
Ho$_x$Y$_{1-x}$PtIn system ($x~\approx~0.04$), for the applied
field continuously rotated within the $(0~0~1)$ or the $(1~2~0)$
plane. The above R = Ho system was preferred because the $M(H)$
curves in Fig.\ref{F16} are consistent with CL-SPM saturated state
at $H~=~55$ kG for both $H~\parallel~c$ and $H~\perp~c$, whereas
the system with low concentration of magnetic ions was chosen for
this measurement in order to check the validity of our model in
the single-ion limit. Moreover, the anisotropic field dependent
data for the diluted sample (lines, Fig.\ref{F16}) show almost
horizontal plateaus for fields higher than $\sim~30$ kG, with
magnetization values close to the corresponding ones for the pure
HoPtIn (symbols). The angle $\alpha$ for Ho$_x$Y$_{1-x}$PtIn,
calculated using $M([0~0~1])~|_{55~kG}~=~7.41~\mu_B/$ Ho, is
$42.1^0$, close to the corresponding value for the pure compound.

\begin{figure}
\begin{center}
\includegraphics[angle=0,width=120mm]{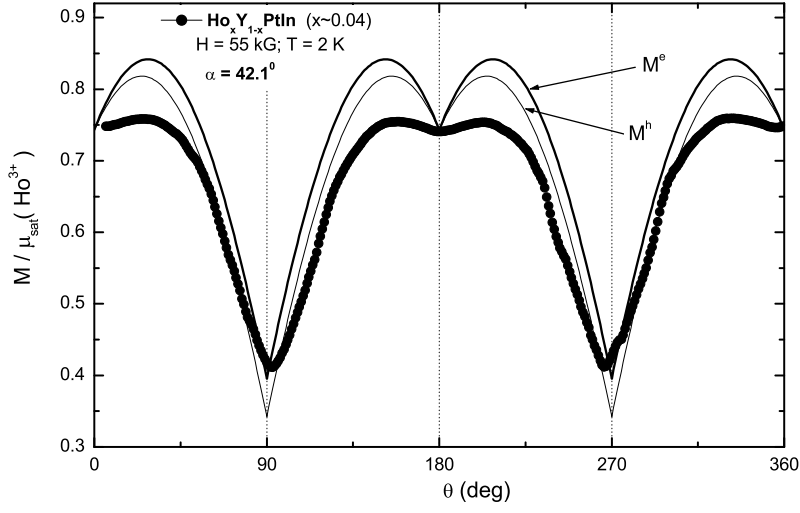}
\end{center}
\caption{Angular dependent magnetization for Ho$_x$Y$_{1-x}$PtIn
(x $\sim~0.04$) (full circles) at H = 55 kG and T = 2 K. The solid
lines represent the "easy" and "hard" plane calculated
magnetizations as a function of $\theta$ (see text), for fixed
angle $\alpha$.}\label{F28}
\end{figure}

The magnetization measured for field applied within the basal
plane (\textit{i.e.,} the (0 0 1) plane) reveals the six-fold
anisotropic data, with the ratio $M([1~1~0])~/~M([1~2~0])$ close
to $cos~30^0~\simeq~0.9$, as expected based on the proposed model.
The angular dependent magnetization at $H~=~55$ kG is shown in
Fig.\ref{F28} (full circles) for $H~\parallel~(1~2~0)$. Also shown
as solid lines are the calculated $M^e(\theta)$ and $M^h(\theta)$,
for fixed $\alpha~=~42.1^0$ determined above. As can be seen, the
measured data qualitatively reproduces the features expected based
on the above model (\textit{i.e.}, two-fold symmetry with respect
to both the \textit{c}-axis and the \textit{ab}-plane, local
minima corresponding to $H~\parallel~[001]$ or $\theta~=~n*180^0$,
n-integer, and maxima at some intermediate angle). More detailed
models which would characterize the RPtIn may exist, and
determining all of them is beyond the scope of this paper.
However, if we restrict our discussion to the three-dimensional
model described before, we see that significant departures from
both calculated $M^e(\theta)$ and $M^h(\theta)$ curves can still
be observed, despite the apparent qualitative agreement between
calculations and measured data. This may mean that either a totaly
different model needs to be considered, or that the aforementioned
model needs further refinement in order to describe at least the
HoPtIn, and perhaps the rest of the RPtIn compounds. Furthermore,
additional experiments (\textit{i.e.}, neutron diffraction) are
required to help identify the most appropriate model for the
magnetization of the RPtIn compounds.

\section{conclusions}

Single crystals of the RPtIn compounds (R = Gd - Lu) have been
grown using the self-flux technique, and have been characterized
by anisotropic temperature- and field-dependent magnetization and
zero-field specific heat measurements. A small Pt-deficiency in
the YbPtIn is apparent from single crystal X-ray data, whereas all
the other heavy R members of the series are believed to form
stoichiometrically. Because of this difference in composition, we
leave the characterization of the YbPtIn system to a separate
study\cite{mor21}, currently underway.

The magnetic RPtIn compounds order magnetically above 2 K, with
the ordering temperatures scaling well with the deGennes dG factor
(Fig.\ref{F25}). This is consistent with the coupling between the
conduction electrons and the local magnetic moments giving rise to
the long-range magnetic order via RKKY exchange interaction.
However, the R = Tb and Tm members of the series have
antiferromagnetic ground states, whereas in the ordered state, the
magnetization of all the other compounds has at least a
ferromagnetic component along the \textit{c}-axis. These
discontinuous changes from antiferromagnetic to ferromagnetic
state across the series seems to also be associated with a change
of low-temperature anisotropy of the paramagnetic state, such that
$\chi_{ab}~>~\chi_c$ for TbPtIn and TmPtIn, and
$\chi_{ab}~<~\chi_c$ for the rest of the magnetic RPtIn.

The magnetization of the TbPtIn compound is extremely anisotropic,
with the magnetic moments confined to the \textit{ab}-plane. Below
the antiferromagnetic ordering temperature $T_N~=~46.0$ K, a
second magnetic phase transition is apparent around 27 K. At low
temperature, in-plane magnetization data reveals complex
metamagnetism, and this has been studied in detail, and described
using the \textit{three co-planar Ising-like systems model} by
Morosan \textit{et al.}\cite{mor14}.

Having understood the complex angular dependent metamagnetism in
the planar TbPtIn compound, we attempted to generalize the
\textit{three co-planar Ising-like systems model} to three
dimensions, such as to characterize the magnetically ordered state
in the other RPtIn compounds: instead of assuming the moments to
be confined to equivalent co-planar directions, $60^0$ away from
each other, they could be restricted to equivalent directions
within vertical planes rotated by $60^0$ around the
\textit{c}-axis. This is equivalent with sets of six non-planar
easy axes, each at an angle $\alpha$ from the \textit{c}-axis,
with "up" and "down" orientations for each directions. When the
applied field is oriented at a non-zero angle from the
\textit{c}-axis, the three magnetic moments will align along the
three easy axes closest to the direction of the field. (This
implies that at high enough fields, all three moments will be in
the "up" positions of three adjacent easy axes, corresponding to
the CL-SPM moment configuration.)

The angle $\alpha$ between the easy axes and the \textit{c}
direction is dependent, in each compound, on the crystalline
electric field CEF energy. Simple geometrical relations allow us
to calculated the expected components of the CL-SPM magnetization
along the \textit{c}-axis, as well as for the "easy" and "hard"
in-plane orientations of the field. Assuming that for $H~=~55$ kG
(in most cases the maximum available field for our measurements),
the RPtIn systems indeed reach the CL-SPM state at low
temperatures, we can determine the fixed value for the angle
$\alpha$ for each compound, and compare the high-field measured
magnetization values with the calculated ones.

As can be seen from Table 2, all RPtIn (R = Tb - Tm) are well
described by this model, with $\alpha$ values between $89^0$ for R
= Tb, and $\sim~32^0$ for R = Er. However, such a model does not
fully account for the angular dependence of the magnetization, at
least in the case of Ho$_x$Y$_{1-x}$PtIn: this is qualitatively
reproduced by the model calculations, with considerable
differences between the measured and theoretical magnetization
values for the whole angular range. Whereas reasonable
misorientation of the rotation sample cannot account for these
differences, we are led to believe that it is necessary to refine
the over-simplified model described here, and also that additional
measurements may help clarify the magnetic structure in these
RPtIn compounds.

\section{appendix}

In a cartesian coordinate system as shown in Fig.\ref{F27}, the
three magnetization vectors in the CL-SPM state can be written as

$\overrightarrow{M}_1~=~\mu_{sat}($R$^{3+})~(0,~\sin\alpha,~\cos\alpha)$,

$\overrightarrow{M}_2=\mu_{sat}($R$^{3+})~(\sin\alpha*\cos30^0,\sin\alpha*\sin30^0,\cos\alpha)$

and

$\overrightarrow{M}_3=\mu_{sat}($R$^{3+})~(\sin\alpha*\cos30^0,-\sin\alpha*\sin30^0,\cos\alpha)$,

whereas, in general, the applied field vector can be written as

$\overrightarrow{H}~=~(H_x,~H_y,~H_z)$.

Thus the general expression for the CL-SPM magnetization M becomes

$M~=~\frac{1}{3}~[(0~+~\sin\alpha*\cos30^0~+~\sin\alpha*\cos30^0)*\frac{H_x}{H}+~
(\sin\alpha~+~\sin\alpha*\sin30^0~-~\sin\alpha*\sin30^0)*\frac{H_y}{H}+\\~
(\cos\alpha~+~\cos\alpha~+~\cos\alpha)*\frac{H_z}{H}]$

or

$M~=~\frac{\sqrt{3}}{3}~\sin\alpha~*\frac{H_x}{H}~+~\frac{1}{3}~\sin\alpha*\frac{H_y}{H}~+\\\cos\alpha*\frac{H_z}{H}$.

Experimentally, we are able to measure the angular dependence of
the magnetization within the "easy" and "hard" planes. If the
magnetic field is continuously rotated within the "easy" plane ((2
1 0) in Fig.\ref{F27}) than, in cartesian coordinates, the vector
$\overrightarrow{H}$ becomes

$\overrightarrow{H}~=~H~(\cos30^0*\sin\theta,~
\sin30^0*\sin\theta,~\cos\theta)$,

where $\theta$ is a continuous variable representing the angle
between the applied field and the \textit{c}-axis.

In this case, the angular dependent magnetization becomes

$M^e/\mu_{sat}($R$^{3+})~=~\frac{\sqrt{3}}{3}~\sin\alpha*\frac{\sqrt{3}}{2}~\sin\theta~$

$+\frac{1}{3}~\sin\alpha*~\frac{1}{2}~\sin\theta~+~\cos\alpha*~\cos\theta~=$

$\frac{2}{3}*\sin\alpha*~\sin\theta~+~\cos\alpha*~\cos\theta$,

where the index "e" refers to the "easy" plane component.

Similarly, if the magnetic field is rotated within the "hard" (1 1
0) plane, the vector $\overrightarrow{H}$ can be written as

$\overrightarrow{H}~=~H~(\cos60^0*\sin\theta,~
\sin60^0*\sin\theta,~\cos\theta))$

and the corresponding angular dependent magnetization is

$M^h/\mu_{sat}($R$^{3+})~=~\frac{\sqrt{3}}{3}~\sin\alpha*\frac{1}{2}~\sin\theta~$

$+\frac{1}{3}~\sin\alpha*~\frac{\sqrt{3}}{2}~\sin\theta~+~\cos\alpha*~\cos\theta~=$

$\frac{\sqrt{3}}{3}*\sin\alpha*~\sin\theta~+~\cos\alpha*~\cos\theta$.

The index "h" is used to indicate the "hard" plane component of
this magnetization.

Both calculated $M^e(\theta)$ and $M^h(\theta)$ are shown in
Fig.\ref{F28} (solid lines) for fixed $\alpha~=~42.1^0$, as
calculated for the Ho$_x$Y$_{1-x}$PtIn system (see text). As
expected, the two-fold symmetry with respect to the
\textit{c}-axis ($\theta~=~2n*90^0$, n-integer) and the
\textit{ab}-plane ($\theta~=~(2n+1)*90^0$, n-integer) is revealed
by both the calculated angular dependent magnetizations.

\section{acknowledgments}

We are thankful to H. B. Rhee, J. D. Strand and S. A. Law for
growing some of the compounds, to Y.A. Mozharivskyj for the single
crystal x-ray measurements, and also to Prof. B. Harmon for
helpful discussions. Ames Laboratory is operated for the U.S.
Department of Energy by Iowa State University under Contract No.
W-7405-Eng.-82. This work was supported by the Director for Energy
Research, Office of Basic Energy Sciences.

\clearpage
\begin{table}
 \begin{center}
 \caption{Magnetic ordering temperatures, $T_m$, effective magnetic
moments and anisotropic paramagnetic Weiss temperatures
$\Theta_W$.} \label{t1}
\begin{tabular}{|c|c|c|c|c|c|c|c|} \hline
$~$&$~$&$~$&$~$&$~$&$~$&$~$&$~$ \\
& Gd & Tb & Dy & Ho & Er & Tm & Yb \\
$~$&$~$&$~$&$~$&$~$&$~$&$~$&$~$ \\
\hline $~$&$~$&$~$&$~$&$~$&$~$&$~$&$~$ \\
T$_m ($K)&$~67.5~\pm~0.5~$ & $~46.0~\pm~0.5~$ &
$~26.5~\pm~0.5~$ & $~23.5~\pm~0.5~$ & $~8.5~\pm~0.5~$ & $~3.0~\pm~0.5~$ & $~2.1~\pm~0.2~$ \\
&  & $27.4~\pm~0.5$ & & & & & \\
$~$&$~$&$~$&$~$&$~$&$~$&$~$&$~$ \\
$\mu_{eff}(\mu_B$) & 7.6 & 9.7 & 10.7 & 10.5 & 10.1 & 7.7 & 4.3 \\
$~$&$~$&$~$&$~$&$~$&$~$&$~$&$~$ \\
$\Theta_{ab}$(K) & $~-57.2~\pm~1.5~$ & $~-38.1~\pm~1.4~$ &
$~-2.2~\pm~5.2~$ & $~7.5~\pm~0.2~$ & $~14.5~\pm~10.8~$ &
$~-7.8~\pm~2.2~$ & $~8.2~\pm~0.9~$ \\
$~$&$~$&$~$&$~$&$~$&$~$&$~$&$~$ \\
$\Theta_{c}$(K) & $~-67.9~\pm~0.5~$ & $~-29.2~\pm~3.1~$ &
$~-29.0~\pm~0.4~$ & $~-27.8~\pm~0.3~$ & $~-9.6~\pm~1.9~$ &
$~36.9~\pm~0.5~$ & $~135.9~\pm~4.0~$ \\
$~$&$~$&$~$&$~$&$~$&$~$&$~$&$~$ \\
$\Theta_{ave}$(K) & $~-61.6~\pm~0.8~$ & $~-34.7~\pm~4.6~$ &
$~-9.1~\pm~0.8~$ & $~-7.7~\pm~1.0~$ & $~13.2~\pm~3.3~$ &
$~2.5~\pm~0.5~$ & $~32.5~\pm~2.8~$ \\
$~$&$~$&$~$&$~$&$~$&$~$&$~$&$~$ \\
\hline
\end{tabular}
\end{center}
\end{table}

\begin{table}
 \begin{center}
 \caption{Anisotropic magnetization values measured at $H~=~55$
kG, and the angles $\alpha$ determined as described in the text.}
\label{t2}
\begin{tabular}{|c|c|c|c|c|c|} \hline
$~$&$~$&$~$&$~$&$~$&$~$ \\
& Tb & Dy & Ho & Er & Tm\\
$~$&$~$&$~$&$~$&$~$&$~$ \\
\hline $~$&$~$&$~$&$~$&$~$&$~$ \\
$~M^{exp}([0~0~1])~/~\mu_{sat}($R$^{3+})~$ & 0.03 & 0.69 & 0.78 & 0.83 & 0.33 \\
$~$&$~$&$~$&$~$&$~$&$~$ \\
$\alpha^{(a)}$ & $~88^0~$ & $~46.5^0~$ & $~38.6^0~$ & $~34.0^0~$ & $~71.0^0~$ \\
$~$&$~$&$~$&$~$&$~$&$~$ \\
$~M^{exp}([1~2~0])~/~\mu_{sat}($R$^{3+})~$ & ~~0.68;~0.62$^{(b)}$~~ & 0.50 & 0.42 & 0.28 & 0.63 \\
$~$&$~$&$~$&$~$&$~$&$~$ \\
$\alpha^e;~\alpha^h$ & $90^0;~~90^0~$ & $~~47.4^0;~~51.3^0~$  & $~~39.7^0;~~43.8^0~$ & $~~29.1^0;~~32.7^0~$ & $~~71.0^0;~~73.4^0~$ \\
$~$&$~$&$~$&$~$&$~$&$~$ \\
$\alpha$ & $~89.0^0~\pm~1.0^0~$ & $~48.9^0~\pm~2.4^0~$ &
$~40.6^0~\pm~2.0^0~$ & $~31.6^0~\pm~2.5^0~$ & $~72.2^0~\pm~1.2^0~$ \\
$~$&$~$&$~$&$~$&$~$&$~$ \\
\hline
\end{tabular}
\end{center}
\end{table}

(a) $\alpha$ values determined using $M([0~0~1])\mid_{55~kG}$ (see
text).

(b) In-plane anisotropic magnetization values ($M^e$ and $M^h$)
used to calculate the respective $\alpha$ values for TbPtIn.

\clearpage

\clearpage

\clearpage

\clearpage

\clearpage

\clearpage

\clearpage

\clearpage

\clearpage

\clearpage

\clearpage

\clearpage

\clearpage

\clearpage

\clearpage

\clearpage

\clearpage

\clearpage

\clearpage

\clearpage

\clearpage

\clearpage

\clearpage

\clearpage

\clearpage

\clearpage

\clearpage

\end{document}